\documentclass[useAMS,usenatbib]{mn2e}
\usepackage{psfig}
\usepackage{hhline}
\usepackage{graphicx}
\usepackage{caption}
\usepackage{bm}

\usepackage{times}
\usepackage{amssymb,amsmath}

\newcommand{\RN}[1]{%
  \textup{\uppercase\expandafter{\romannumeral#1}}%
}

\def\lsim{ \lower .75ex\hbox{$\sim$} \llap{\raise .27ex \hbox{$<$}} }
\def\gsim{ \lower .75ex \hbox{$\sim$} \llap{\raise .27ex \hbox{$>$}} }

\title[Breaking degeneracy in jet dynamics: multi-epoch joint modelling of the BL Lac PKS\,2155$-$304]{Breaking degeneracy in jet dynamics: multi-epoch joint modelling of the BL Lac PKS\,2155$-$304}

\author[M. Lucchini, S. Markoff, P. Crumley, F. Krau{\ss}, R. M. T. Connors]
{M. Lucchini$^1$\thanks{E--mail: m.lucchini@uva.nl}, S. Markoff$^{1,2}$, P. Crumley$^{1,2,3}$, F. Krau{\ss}$^{1,2}$, R. M. T. Connors$^{1,4}$\\
$^1$API -- Anton Pannekoek Institute for Astronomy, University of Amsterdam, Science Park 904, 1098 XH Amsterdam, the Netherlands\\
$^2$GRAPPA -- Gravitational and Astroparticle Physics Amsterdam, University of Amsterdam, Science Park 904, 1098 XH Amsterdam, the Netherlands\\
$^3$Department of Astrophysical Sciences, Princeton University, 4 Ivy Lane, NJ 08544 Princeton, New Jersey, USA\\
$^4$Cahill Center for Astronomy and Astrophysics, California Institute of Technology, 1200 California Boulevard, Pasadena, CA 91125, USA\\
}

\begin{document}

\maketitle

\begin{abstract} 
Supermassive black holes can launch powerful jets which can be some of the most luminous multi-wavelength sources; decades after their discovery their physics and energetics are still poorly understood. The past decade has seen a dramatic improvement in the quality of available data, but despite this improvement the semi-analytical modelling of jets has advanced slowly: simple one-zone models are still the most commonly employed method of interpreting data, in particular for AGN jets. These models can roughly constrain the properties of jets but they cannot unambiguously couple their emission to the launching regions and internal dynamics, which can be probed with simulations. However, simulations are not easily comparable to observations because they cannot yet self-consistently predict spectra. We present an advanced semi-analytical model which accounts for the dynamics of the whole jet, starting from a simplified parametrization of Relativistic Magnetohydrodynamics in which the magnetic flux is converted into bulk kinetic energy. To benchmark the model we fit six quasi-simultaneous, multi-wavelength spectral energy distributions of the BL Lac PKS\,2155$-$304 obtained by the TANAMI program, and we address the degeneracies inherent to such a complex model by employing a state-of-the-art exploration of parameter space, which so far has been mostly neglected in the study of AGN jets. We find that this new approach is much more effective than a single-epoch fit in providing meaningful constraints on model parameters.
\end{abstract}


\section{Introduction}
Accreting compact objects such as neutron stars and black holes often display collimated and relativistic outflows called jets. Jets have been observed (among others) in X-ray binaries (e.g. \citealt{Mirabel94}, \citealt{Fender97}) and in active galactic nuclei (e.g. \citealt{Fanaroff74}) ; their emission can span many orders of magnitude in frequency, from radio up to TeV $\gamma$-ray. They appear to be more common when the accretion rate on the compact object is either very sub-Eddington (roughly below $1\%$ of the Eddington luminosity), as is the case in hard state X-ray binaries and low-luminosity AGNs (\citealt{Ho08} and references therein, but see \citealt{Ghisellini14}) or highly super-Eddington, as in Gamma-ray Bursts (e.g. \citealt{Sari99}) and jetted Tidal Disruption Events (\citealt{Bloom11}, \citealt{Burrows11}, \citealt{Cenko12}). 

Despite their prevalence, jets are still poorly understood astrophysical sources. In the standard picture, black hole jets are collimated and launched by magnetic fields dragged near the event horizon and wound up either through frame dragging caused by a rotating black hole \citep{Blandford77}, or by differential rotation in the accretion disk \citep{Blandford82}. At some distance from the central engine, particles are accelerated continuously to relativistic energies by internal shocks within the jet \citep{Blandford79}.

On the theoretical side, in the past fifteen years General Relativistic Magnetohydrodynamics (GRMHD) simulations have made great strides in demonstrating how jets are powered, launched, and accelerated (e.g. \citealt{Gammie03}, \citealt{DeVilliers05}, \citealt{McKinney06}, \citealt{Tchekhovskoy11}, \citealt{Liska18}). However, numerical GRMHD codes do not yet self-consistently handle radiation/radiation transfer and/or are extremely computationally intensive. While some efforts have been made to make simulations comparable with observations (e.g. \citealt{Monika09}, \citealt{Monika17}, \citealt{Dai18}) these efforts are still in their early days. Furthermore, GRMHD codes typically assume an ideal, single temperature fluid and do not yet have the dynamic range to capture the microscopic scales over which particle acceleration occurs. 

On the phenomenological side, semi-analytical models successfully predict most spectral energy distributions (SEDs) of these objects but are often either overly simplistic or degenerate. 

In recent years, the favoured approach (especially in the case of AGN jets) has been the so-called ``one-zone model'' in which the bulk of the emission comes from a single spherical region, often close to the jet base (e.g. \citealt{Tavecchio98}). While these models can usually reproduce SEDs fairly well, the typical synchrotron self-absorption frequency in these models is of the order of $10^{11-12}\,\rm{Hz}$, and thus the radio emission is assumed to originate in regions of the jet further away from the black hole. Therefore, one-zone models cannot unambiguously couple the observed emission to the launching regions and internal jet dynamics. Furthermore, the erratic variability observed in many jetted AGN in both timing (e.g. \citealt{Aharonian07}, \citealt{Acciari11}) and polarimetry (e.g. \citealt{Marscher08}, \citealt{Blinov18}) also calls for physics that can't be captured in a homogeneous single-zone model.

Structured models which account for the emission of the entire outflow actually pre-date one zone models (e.g. \citealt{Marscher80}, \citealt{Konigl81}, \citealt{Ghisellini85}), but they have fallen out of favour with the advent of modern observational facilities, particularly due to the discovery that the high-energy emission in many AGN can vary on extremely short time-scales, implying a small size of the emitting region (e.g. \citealt{Aharonian07}). However, in recent times the simplest one-zone model approach has been called into question due to both the erratic variability detected in some AGN (e. g. \citealt{Bottcher10b}) or the extreme parameters required by some sources \citep{Tavecchio16}. These issues, together with a desire to model radio fluxes and probe jet morphologies and dynamics, have made inhomogeneous multi-zone model an attractive option once again. Modern multi-zone models generally come in one of three flavours: structured outflows that invoke Doppler boosting between different regions, such as a fast-moving spine and a slow layer (e.g. \citealt{Ghisellini05}) or a decelerating jet \citep{Georganopoulos03}, detailed shock-in-jet models that focus on electron dynamics (e.g. \citealt{Bottcher10a}, \citealt{Malzac13}), and extended outflow models whose aim is to capture the dynamics and/or energetics of the entire jet, rather than focusing on a single region (e.g. \citealt{Markoff05}, \citealt{Boutelier08}, \citealt{Potter13a}). Unfortunately, moving away from the single-zone paradigm introduces large numbers of free parameters and/or increases the computational cost (particularly if one aims at also predicting variability or polarization signatures e.g. \citealt{Marscher14}, \citealt{Potter18}), which results in severe model degeneracies and/or in a loss of predictive power. 

The limitations in both simulations and phenomenological models, as well as the lack of robust methods for fitting data statistically, make it paramount to develop more accurate semi-analytical models along with fitting techniques capable of reducing possible degeneracies. 

Further complicating matters, in many low-power sources the contribution of the jet to the SED  is poorly constrained, and other processes such as inverse Comptonization from a corona (e.g. \citealt{Shapiro76}, \citealt{Haardt93}) and/or or the contribution from a radiatively inefficient accretion flow (e.g. \citealt{Narayan96}) can also contribute to the emission, particularly in the X-ray band. In this context, highly beamed sources are a particularly useful tool for isolating jet physics. 

Blazars are ideal sources for this purpose. These are radio-bright AGN with one of the jets pointed towards the observer (e.g. \citealt{Blandford78}, \citealt{Urry95}, and see \citealt{Ghisellini13} for a review). Because of relativistic beaming, the radiation produced in the jet can outshine all other components such as the accretion disc, corona or dusty torus. Blazars are divided into two categories: Flat Spectrum Radio Quasars (FSRQs) and BL Lacertae (BL Lacs), depending on whether they show bright (equivalent width $>5$ $\rm {\AA}$) or faint/absent (equivalent width $<5$ $\rm {\AA}$) emission lines in their optical spectrum. All blazars show two humps in their Spectral Energy Distribution (SED); the first originates from relativistic electrons emitting synchrotron radiation in the jet, while the second is often attributed to Inverse Compton scattering (IC) either with photons produced in the jet (synchrotron-self Compton, SSC), or coming from the external environment (EC, external Compton). Possible target photon fields include the emission from the disc, broadened line emission coming from ionized clouds of gas orbiting close to the black hole (BLR, broad line region), or a dusty torus surrounding the disc. Alternatively, the second hump could be caused by hadronic processes initiated by a population of relativistic protons. 

Typically, efforts to model blazars have been limited to the study of a single multi-wavelength SED for a given source, taken either during organized campaigns, sometimes during flaring states, or by utilizing archival, non-simultaneous data. This greatly limits the ability of any model to effectively constrain the physics of the source. Only recently has it been possible to compile multi-wavelength, multi-epoch, \textit{and} quasi-simultaneous SEDs, for example thorough the TANAMI\footnote{HTTPS://fekrauss.com/resources/} (Tracking Active Galactic Nuclei with Austral Milliarc-second Interferometry) multi-wavelength program (\citealt{Ohja10}, \citealt{Krauss16}).

In this work we model six quasi-simultaneous, radio through $\gamma$--ray SEDs of the BL Lac PKS\,2155$-$304 obtained during the TANAMI campaign with a new steady-state, multi-zone, semi-analytical dynamical model designed as a simple parametrization of relativistic MHD. The treatment of particle acceleration and radiation are identical to those of the \texttt{agnjet} model, developed by \citealt{Markoff05}, \citealt{Maitra09}, \citealt{Connors17} and including the modifications described in \citealt{Connors18} (submitted). \texttt{Agnjet} has mainly been used to study black hole X--ray binaries in the hard state and in quiescence (e.g. \citealt{Markoff05}, \citealt{Gallo07}, \citealt{Maitra09}, \citealt{Plotkin15}, \citealt{Connors17}). It can also reproduce the broadband SEDs of LLAGN (e.g. Sgr A*: \citealt{Falcke00}, \citealt{Markoff01} \citealt{Connors17}; M81: \citealt{Markoff08}, \citealt{Markoff15}; NGC 4051: \citealt{Maitra11}) and of the nearby low power FR\RN{1} radio galaxy M87 (\citealt{Prieto16}); however, in \texttt{agnjet} the outflow velocity is limited to low bulk Lorentz factors, making it incapable of treating more powerful blazar jets \citep{Crumley17}. The new dynamical model described in this paper is called \texttt{bljet}. Furthermore, we combine our new dynamical model with a thorough exploration of parameter space, and show that, compared to individual SEDs, multi-epoch data can provide much stronger constraints on model parameters.

The paper is structured as follows: in Section 2 we present the updated \texttt{bljet} model, in Section 3 we use it to fit six quasi--simultaneous SEDs of PKS\,2155$-$304 from the TANAMI program, and compare individual and joint fits of the data, in Section 4 we discuss our results, and in Section 5 we draw our conclusions.

Throughout the paper we use cgs units and assume the following cosmological parameters:
$H_0=69.6$ km s$^{-1}$ Mpc$^{-1}$, $\Omega_{\rm M}=0.286$, $\Omega_{\Lambda}=0.714$ \citep{Bennett14}. With this choice, the luminosity distance of PKS\,2155$-$304, located at redshift $z=0.116$, is 543.4 Mpc.

\begin{figure}
\hspace{-0.5cm}
\includegraphics[scale=0.45]{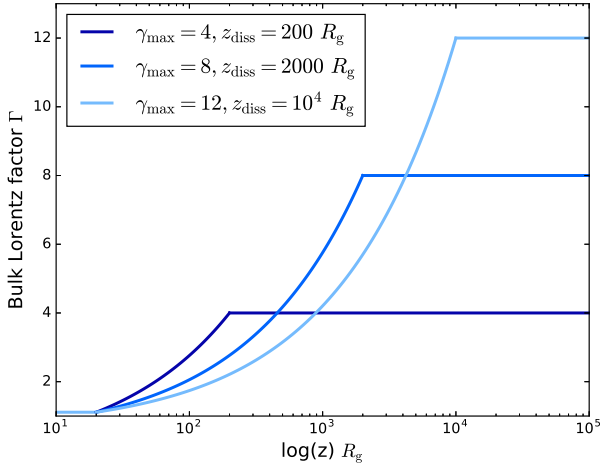}
\caption{
Jet bulk Lorentz factor as a function of distance $z$ from the black hole, taking $z_{0} = 20$ $R_{\rm g}$, $\gamma_{\rm max} = 4$, $8$, and $12$, and $z_{\rm acc} = 200$, $2000$, and $10^{4}$ $R_{\rm g}$.
}
\label{gammap}
\end{figure}

\section{Model description}
In order to describe the properties of a given jet it is necessary to know its velocity (or velocity profile, in the case of multi--zone models such as ours), as well as its energy budget and the way this is divided between proton and electron internal and kinetic energy, and magnetic fields. The goal of this section is to derive some simple analytical expressions for these quantities, which can then be used to produce SEDs to be compared with observations. We start with the velocity profile and magnetic field.

We assume that a certain amount of power, expressed as a percentage $N_{\rm j}$ of the black hole's Eddington luminosity $L_{\rm Edd}$, is channelled from the inner radius of the accretion disc into an outflowing cylinder situated above the disc, with radius $r_0$ (measured in units of $R_{\rm g}$) and up to a height $z_0 = h \cdot r_0$, which we term the nozzle of the jet. This initial power is divided between particles (electrons and protons) and a magnetic field. The nozzle represents the base of the jet and may correspond to an outflowing, lamp-post corona \citep{Martocchia96}. Unlike in \texttt{agnjet}, the jet is always magnetically dominated near the base, and the initial magnetic field is assumed to be converted into bulk kinetic energy through acceleration of the outflow. Because we cannot treat jet acceleration in full GRMHD, we parametrize this behaviour with a special relativistic prescription in which energy is conserved.

\begin{figure}
\hspace{-0.5cm}
\includegraphics[scale=0.45]{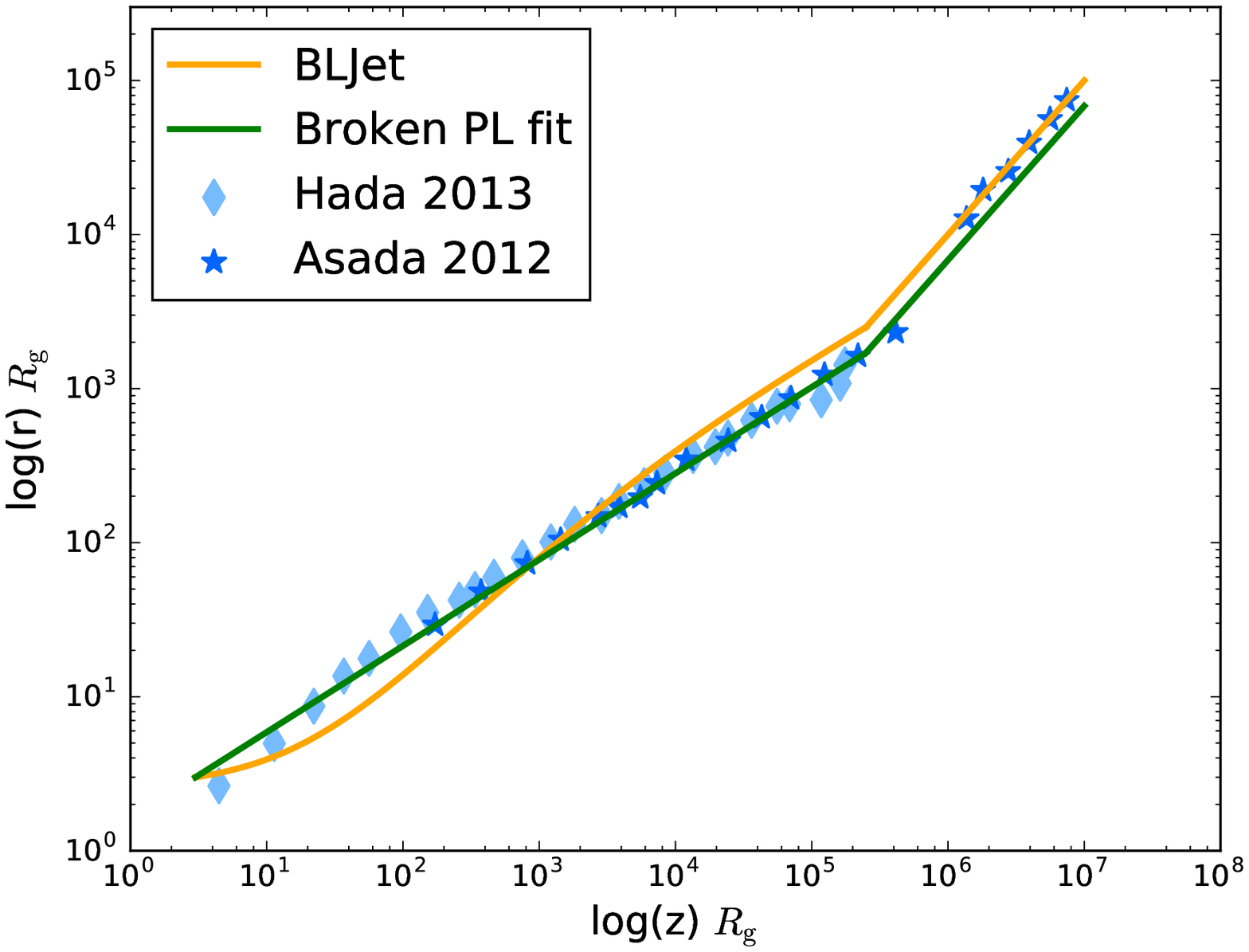}
\caption{
Jet radius as a function of distance from the black hole for the jet in M87. Stars show VLBI data reported in \citealt{Asada12}, diamonds show data from \citealt{Hada13}. The green line shows the broken power-law fit; the inner region is parabolic ($r \propto z^{0.56}$, \citealt{Hada13}) and the outer region is conical ($r \propto z$, \citealt{Asada12}). The orange line is the profile used in \texttt{bljet} with $r_{\rm 0} = 3\,{\rm R_g}$, $h = 2$, $z_{\rm acc} = 2.5\cdot10^{5}\,{\rm R_g}$, $\gamma_{\rm acc} = 15$. 
}
\label{m87}
\end{figure}

We assume that the jet bulk acceleration begins at the top of the nozzle, starting with with an initial Lorentz factor $\gamma_0$ at a height $z_0$, and continues until a final Lorentz factor $\gamma_{\rm acc}$ is achieved at a distance $z_{\rm acc}$. Beyond this region the jet velocity remains unchanged. The initial bulk Lorentz factor is assumed to be $\gamma_{\rm 0} = 1.09$ ($\beta_{\rm 0}= 0.4$), which in the old \texttt{agnjet} model corresponds to the maximal sound speed; $\gamma_{\rm 0}$ has a negligible effect on the SED. The bulk Lorentz factor in the acceleration region is assumed to follow a power-law in the acceleration region of index $\alpha = 1/2$, as suggested by analytical MHD (e.g. \citealt{Vlahakis04}, \citealt{Beskin06}, \citealt{Komissarov07}). Three different possible velocity profiles with differing terminal velocities are shown in Fig.~\ref{gammap}. The region of the jet close to and downstream of $z_{\rm acc}$ can be thought of as the equivalent of the ``blazar zone'' probed by one-zone models, and throughout the paper we use the two terms interchangeably. The velocity profile in the acceleration region is thus: 
\begin{equation}
\gamma(z) = \gamma_0 + \left(\gamma_{\rm acc}-\gamma_0\right)\frac{z^{1/2} - z_{0}^{1/2}}{z_{\rm acc}^{1/2} - z_{0}^{1/2}}.
\label{gamma}
\end{equation}
For every $z$, we take the jet opening angle to be inversely proportional to the Lorentz factor as suggested by VLBI observations of \textit{Fermi}/LAT detected blazars (\citealt{Pushkarev09}, see also \citealt{Jorstad05}, \citealt{Clausen13}, \citealt{Pushkarev17}):
\begin{equation}
\theta(z) = \frac{0.15}{\gamma(z)},
\label{theta}
\end{equation}
where the factor of $0.15$ is a typical value inferred from the same VLBI campaigns. We take the jet radius as a function of distance to be:
\begin{equation}
r(z) = r_0+(z-z_0)\tan(\theta(z)),
\label{r}
\end{equation}
This results in a jet that is roughly parabolic ($r(z) \propto z^{1/2}$) as it accelerates near its base, and which then expands conically ($r(z) \propto z$) after reaching its terminal speed; therefore, our choice of velocity profile results in a radial profile that is roughly consistent with that observed in M87 (\citealt{Asada12}, \citealt{Hada13}, \citealt{Hada16}) as shown in figure \ref{m87}. Unlike the model of \cite{Potter13a}, who assumed a fixed geometry identical to that of M87 for every source, we leave $z_{\rm acc}$ as a free parameter, thus allowing jets with different sizes for the parabolic to conical transition regions. We will show in sections 3 and 4 that this has important consequences for our modelling.  

Particle conservation determines the number density of particles (leptons or protons) along the entire jet to be:
\begin{equation}
n(z) = n_0 \left(\frac{\gamma(z) \beta(z)}{\gamma_0 \beta_0}\right)^{-1}\left(\frac{r(z)}{r_0}\right)^{-2},
\label{particles}
\end{equation}
where $n_0$ is the initial particle number density. We assume a heavy jet containing one cold proton per electron throughout its length: $n(z) = n_{\rm e}(z) = n_{\rm p}(z)$. We choose to restrict ourselves to this regime because as we will show in this section, assuming that a dominant proton contribution is carrying the bulk of the jet's kinetic energy (and enthalpy) greatly simplifies the calculation of the magnetic field strength throughout the jet. In order to conserve energy while accelerating, a streamline of the jet must satisfy the relativistic Bernoulli equation (\citealt{Konigl80}):
\begin{equation}
\gamma(z)\frac{\omega(z)}{n(z)} = \rm{const},
\label{Bernoulli}
\end{equation}
where
\begin{equation}
\omega(z) = U_{\rm p}(z) + U_{\rm e}(z) + P_{\rm e}(z) + U_{\rm b}(z) + P_{\rm b}(z) 
\label{enth}
\end{equation}
\begin{equation*}
= n(z)m_{\rm p}c^{2} + n(z)\langle\gamma_{\rm e}\rangle m_{\rm e}c^{2} + P_{\rm e}(z) + U_{\rm b}(z) + P_{\rm b}(z),
\end{equation*}

\begin{figure}
\hspace{-0.5cm}
\includegraphics[scale=0.45]{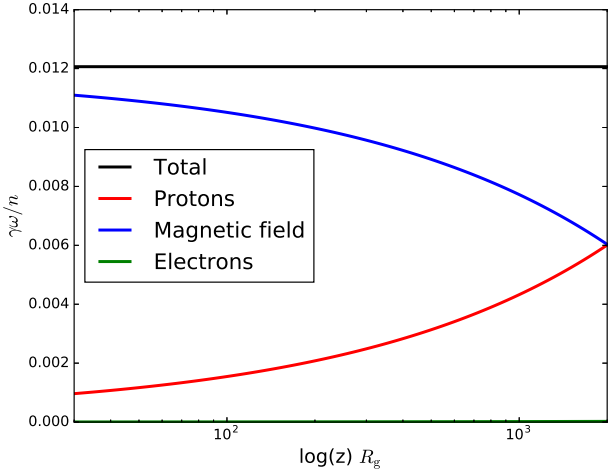}
\caption{
Bernoulli's equation evaluated at each point in the jet acceleration zone, for the same parameters shown so far ($\gamma_{\rm acc} = 8$, $z_{0} = 20$\,$R_{\rm g}$, $z_{\rm acc} = 2000$\,$R_{\rm g}$, $T_e = 10^{10}$\,$K$, $\sigma_{0} = 13.4$, $\sigma_{\rm diss} = 1$, and $N_{\rm j} = 1.38 \times 10^{45}$\,ergs\,s$^{-1}$). The blue, red and green lines represent the the contributions by the magnetic field, protons and electrons respectively; the black solid line shows the sum of all three contributions. As long as the electron contribution is negligible, the total energy is roughly conserved.
}
\label{energy}
\end{figure}

is the total enthalpy of the jet, assuming that the protons are cold and therefore their pressure is negligible. $\langle\gamma_e\rangle$ is the average Lorentz factor of the electrons, $U_{\rm p}(z)$ is the energy density of the cold protons, $U_{\rm e}(z)$ and $P_{\rm e}(z)$ are the internal energy and pressure of the electrons, and $U_{\rm e}(z)$ and $P_{\rm e}(z)$ those of the magnetic field. Because $U_{\rm b}(z)=P_{\rm b}(z)=B^{2}(z)/8\pi$, writing the first three terms explicitly and combining them with Eq.~\ref{gamma}, \ref{theta}, \ref{r}, \ref{particles} and \ref{Bernoulli} allows us to compute the magnetic field profile required to accelerate the jet. The electron pressure and internal energy can be related to each other via the adiabatic index $\Gamma$:
\begin{equation}
P_{\rm e} = \left(\Gamma-1\right) U_{\rm e};
\end{equation}
We only consider relativistic leptons, and therefore $\Gamma = 4/3$. Eq.~\ref{enth} can be written as:
\begin{equation}
\omega(z) = n(z)m_{\rm p}c^{2} + \omega_{\rm e}(z) + B^{2}(z)/4\pi,
\label{enth2}
\end{equation}
where $\omega_{\rm e}(z) = \Gamma U_{\rm e}(z) = \Gamma  n(z) \langle\gamma_e\rangle m_{\rm e} c^{2}$ is the electron contribution to the total enthalpy.  We define the magnetization parameter $\sigma$ as:
\begin{equation}
\sigma(z) = \frac{P_{\rm b}(z) + U_{\rm b}(z)}{U_{\rm p}(z)+ U_{\rm e}(z) + P_{\rm e}(z)} 
\label{sigma}
\end{equation}
\begin{equation*}
= \frac{B^{2}(z)}{4\pi\left(n(z)m_p c^{2} + \Gamma n(z) \langle\gamma_e\rangle m_{e} c^{2}\right)};
\end{equation*}
Eq.~\ref{sigma} allows us to write Eq.~\ref{enth2} as:
\begin{equation}
\omega(z) = \left[n(z)m_{\rm p}c^{2} + \omega_{\rm e}(z)\right]\left[1+\sigma(z)\right].
\label{enth3}
\end{equation}

We leave the magnetization at the end of the acceleration region $\sigma_{\rm diss}$ as a free parameter in the model. Because the Bernoulli equation holds at every distance along a streamline of the jet (as long as energy is conserved), we can evaluate it at $z_{0}$ and $z_{\rm acc}$ to find the initial magnetization necessary to reach a desired final Lorentz factor $\gamma_{\rm acc}$:
\begin{equation}
\gamma(z_{\rm acc})\frac{\omega(z_{\rm acc})}{n(z_{\rm acc})} = \gamma(z_{0})\frac{\omega(z_{0})}{n(z_{0})},
\label{Bern0}
\end{equation}
which can be written as
\begin{equation}
\gamma_{\rm acc} \left(1+\sigma_{\rm diss} \right)\left(m_{\rm p} c^{2}+\Gamma \langle\gamma_e\rangle m_{\rm e} c^{2}\right)
\label{Bern1}
\end{equation}
\begin{equation*}
= \gamma_0\left(1+\sigma_0 \right)\left(m_{\rm p} c^{2}+\Gamma \langle\gamma_e\rangle m_{\rm e} c^{2}\right).
\end{equation*}
We only consider isothermal, cold jets in which the average Lorentz factor of the electrons is low $\left(\langle\gamma_{\rm e}\rangle\lesssim 50\right)$ and constant up to $z_{\rm diss}$, so that the proton contribution to the total enthalpy is always much greater than that of the electrons. This assumption also implies that the energy required by any mechanism to offset adiabatic losses \citep{Blandford79} is negligible with respect to the total energy budget. In this regime, the Bernoulli equation simplifies to:
\begin{equation}
\gamma(z_{\rm acc})\left(1+\sigma_{\rm diss}\right) = \gamma_0\left(1+\sigma_0\right),
\label{Bern2}
\end{equation}
so that the required magnetization at the base, as a function of the final magnetization and bulk Lorentz factor, is:
\begin{equation}
\sigma_0 = \left(1+\sigma_{\rm diss} \right) \frac{\gamma_{\rm acc}}{\gamma_0}-1.
\label{sig0}
\end{equation}

We can now evaluate the Bernoulli equation at every $z$ to find the magnetization as a function of distance and jet bulk velocity:
\begin{equation}
\sigma(z) = \frac{\gamma_0}{\gamma(z)}\left(1+\sigma_0 \right)-1
\end{equation}
and by inverting the definition of $\sigma(z)$ we can determine the corresponding magnetic field:
\begin{equation}
B(z) = \left[4\pi \sigma(z) n(z)\left(m_{\rm p} c^{2}+\Gamma\langle\gamma_e\rangle m_{\rm e} c^{2}\right)\right]^{1/2}.
\label{B}
\end{equation}

\begin{figure}
\hspace{-0.5cm}
\includegraphics[scale=0.45]{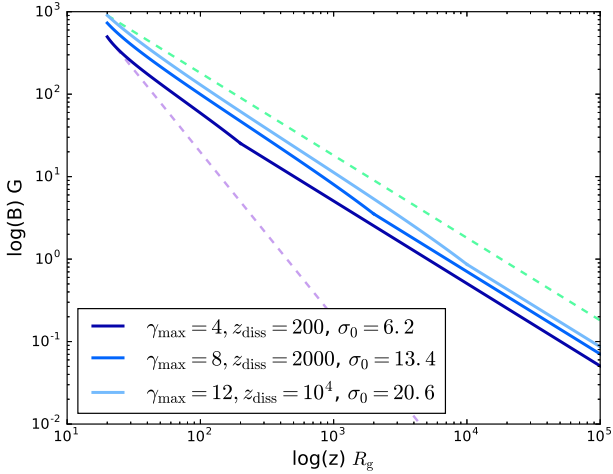}
\caption{
The blue lines represent our analytical solution for the magnetic field as a function of distance $z$ from the black hole in the jet acceleration zone, for the same parameters as Fig.~\ref{gammap}, assuming that the electrons are isothermal with $T_e = 10^{10}$\,$K$ and that $\sigma_{\rm diss} = 1$. The total jet power is $10^{-2}L_{\rm Edd}$, which corresponds to $1.38\times 10^{45}$\,ergs\,s$^{-1}$ for a black hole mass of $M_{\rm bh} = 10^{9}$\,$M_{\odot}$. The dashed green and purple lines represent reference toroidal ($\propto z^{-1}$) and poloidal ($\propto z^{-2}$) fields, respectively. As the terminal Lorentz factor increases, so does the initial magnetization $\sigma_0$.
}
\label{bfield}
\end{figure}

We stress that this result is only valid as long as the second
term in the parentheses is much smaller than the first, meaning that
the protons have to carry the bulk of the particle energy budget. If
this isn’t the case, then our simplification between Eq. \ref{Bern1} and \ref{Bern2} is incorrect. If instead the energy density of the electrons (or leptons for a pair-dominated jet) dominates, our solution would need to account accurately both for adiabatic cooling and the conversion of bulk kinetic or magnetic energy into internal energy of the electrons through either shocks or reconnection, which would impact the values of $\sigma(z)$, $\gamma(z)$ and $\langle \gamma_e(z)\rangle$. The full description of these effects is beyond the scope of this work and will be investigated in a future paper. As long as the jet is cold our solution of the Bernoulli equation always holds and therefore the jet conserves energy while it accelerates, as shown in Fig.~\ref{energy}. 

Beyond the jet acceleration zone the magnetic field is assumed to be purely toroidal:
\begin{equation}
B(z) = B(z_{\rm acc})\frac{z_{\rm acc}}{z},
\end{equation}
in order to reproduce the flat radio spectrum observed in most compact jets, assuming the radiating particle distribution is isothermal (\citealt{Blandford79}). Fig.~\ref{bfield} shows three possible solutions for the magnetic field.

We now need to calculate the initial number density of particles in the jet. The energy density at the base of the jet can be written as:
\begin{equation}
U_{\rm j}(z_0) = U_{\rm e}(z_0)+ U_{\rm p}(z_0) + U_{\rm b}(z_0) = \frac{N_{\rm j}L_{\rm Edd}}{2\pi r_0^{2}\gamma_0 \beta_0 c},
\label{jetrat}
\end{equation}
where the factor 2 accounts for the launching of a jet and a counter-jet, and $N_{\rm j}$ is the total power injected in the jet in Eddington units. We define the standard plasma-beta parameter at the base of the jet as:
\begin{equation}
\beta_{\rm p,0} = \frac{U_{\rm e}(z_0)}{U_{\rm b}(z_0)};
\label{betapl}
\end{equation} 
whose value is set by our assumption of $n_{\rm e}=n_{\rm p}$ and by the initial magnetization defined in Eq.\ref{sig0}:
\begin{equation}
\sigma_{\rm 0} = \frac{2U_{\rm b}(z_{\rm 0})}{U_{\rm p}(z_{\rm 0})+\Gamma U_{\rm e}(z_{\rm 0})} = \frac{2U_{\rm e}(z_{\rm 0})}{\beta_{\rm p,0}\left(U_{\rm p}(z_{\rm 0})+\Gamma U_{\rm e}(z_{\rm 0})\right)},
\end{equation}
from which we find:
\begin{equation}
\beta_{\rm p,0} = \frac{2\langle \gamma_{\rm e} \rangle m_{\rm e}}{\sigma_{\rm 0}\left(m_{\rm p}+\Gamma\langle \gamma_{\rm e} \rangle m_{\rm e}\right)}.
\end{equation}
We can now calculate the initial particle number density from Eq.~\ref{jetrat} and Eq.~\ref{betapl}:
\begin{equation}
n_{\rm 0} = \frac{N_{\rm j}L_{\rm Edd}}{2\pi r_0^{2} \gamma_0 \beta_0 c} \cdot \frac{1}{m_{\rm p} c^{2}+ \langle \gamma_{\rm e} \rangle m_{\rm e} c^{2} \left(1+1/\beta_{\rm p,0} \right)}.
\end{equation}

The injected electrons are initially described by a relativistic Maxwellian distribution having temperature $T_e$ (corresponding to a scale Lorentz factor $\gamma_{\rm th}$). At a distance $z_{\rm diss}$ from the black hole the jet meets a dissipation region beyond which the leptons are heated, which we parametrise by increasing the peak of the thermal Maxwell-J{\"u}ttner distribution $\gamma_{\rm th}$ by a fixed factor $f_{\rm heat}$, and at the same time a fraction $f_{\rm pl}$ of the total number of electrons are assumed from this point onwards to be continuously accelerated into a non-thermal tail, described by a power-law with index $p$. This roughly mimics the behaviour of shock acceleration seen in PIC simulations (e.g. \citealt{Sironi11}). The parameter $f_{\rm heat}$ therefore effectively sets the minimum Lorentz factor $\gamma_{\rm min}$ of the non-thermal particles, which is assumed to scale with the peak of the Maxwellian distribution:
\begin{equation}
\gamma_{\rm min} = 2.23f_{\rm heat} \gamma_{\rm th}
\label{emin}
\end{equation}

We do not specify the mechanism responsible for particle heating and acceleration beyond the dissipation region; instead, the efficiency of this process is quantified through a free parameter $f_{\rm sc}$, which is used to define the particle acceleration time scale independently of the acceleration mechanism: 
\begin{equation}
t_{\rm acc} = \frac{4 \gamma m_{\rm e}c}{3 f_{\rm sc}eB},
\label{tacc}
\end{equation}
where $e$ is the electron charge, $B$ the magnetic field strength, $\gamma$ the electron's Lorentz factor, $m_{\rm e}$ the mass of the electron, and $c$ the speed of light.  In the case  of standard quasi-parallel shock acceleration, $f_{\rm sc} = \beta_{\rm sh}^{2}/(\lambda/R_{\rm gyro})$ \citep{Jokipii87}, where $\beta_{\rm sh}$ is the shock speed relative to the plasma, $\lambda$ is the scattering mean free path of the particles and $R_{\rm gyro}$ their gyroradius. While we do not assume particles are accelerated in shocks, we do assume that $f_{sc}$ does not depend of energy of the particle. The maximum Lorentz factor reached by the leptons is then set by solving:
\begin{equation}
t_{\rm acc}^{-1} = t_{\rm syn}^{-1} + t_{\rm com}^{-1} + t_{\rm dyn}^{-1},
\label{emax}
\end{equation}
where $t_{\rm syn}$ and $t_{\rm com}$ are the synchrotron and Compton radiative time scales at each point in the jet ($t_{\rm syn/com} = 3m_{\rm e}c^{2} / 4\sigma_{\rm t}c U_{\rm rad} \gamma_{\rm e}$, where $U_{\rm rad}$ is the magnetic or photon energy density for synchrotron/inverse Compton respectively and $\gamma_{\rm e}$ the electron Lorentz factor). We take the dynamical time scale to be $t_{\rm dyn} = f_{\rm b} r(z)/\beta(z) c$, where $f_{\rm b}$ is a free parameter and $r(z)$ is the jet radius defined in eq. \ref{r}. The free parameter $f_{\rm b}$ absorbs the uncertainty in the electron diffusion coefficient and importance of adiabatic losses within the jet, similarly to how $f_{\rm sc}$ absorbs our ignorance of the details of particle acceleration. Because $t_{\rm dyn}\gg t_{\rm syn,\,com} \geq t_{\rm acc}$, the parameter $f_{\rm b} $ has a negligible impact on the maximum lepton energy (but it does impact the break energy, see below). In this way, the maximum energy reached by the non-thermal tail is directly linked to the efficiency of the acceleration mechanism as well as local conditions in the jet at each point.

Similarly, we parametrise the energy of the cooling break in the leptons by balancing the radiative and dynamical time scale in each section of the jet.  For simplicity we only consider synchrotron losses when computing the break energy. In this case:
\begin{equation}
t_{\rm syn} = \frac{3m_{\rm e}^{2}c^{3}}{4\sigma_{\rm t}U_{\rm b}(z) E_{\rm br}(z)} = f_{\rm b}\frac{r(z)}{\beta(z) c} = t_{\rm dyn},
\end{equation} 
from which we find the break energy:
\begin{equation}
E_{\rm br}(z) = \frac{3\beta(z) m_{\rm e}^{2}c^{4}}{4f_{\rm b}  r(z)\sigma_{\rm t}U_{\rm b}(z)}.
\label{ebreak}
\end{equation}

\begin{figure}
\hspace*{-0.5cm}
\includegraphics[scale=.55]{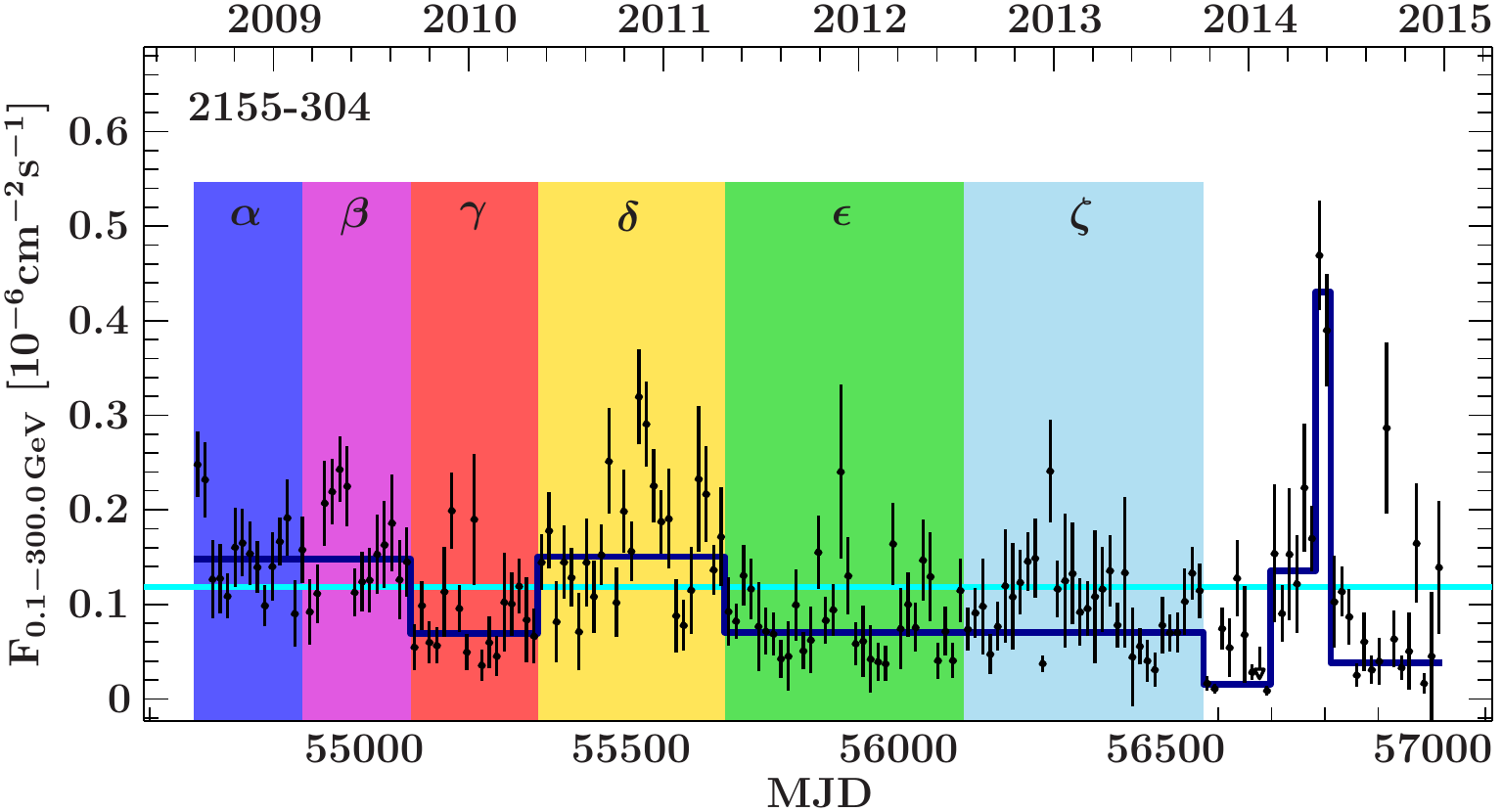}
\caption{
\textit{Fermi}/LAT light-curve of PKS$\,$2155$-$304 from K16. The six epochs $\alpha$, $\beta$, $\gamma$, $\delta$, $\epsilon$, $\zeta$, are highlighted in blue, magenta, red, yellow, green and blue respectively.
}
\label{fermilc}
\end{figure}

In our numerical code $z_{\rm acc}$ and $z_{\rm diss}$ are allowed to have different values; however, in this work we always take $z_{\rm acc} = z_{\rm diss}$, as this is a natural choice and is suggested by observations, e.g. \cite{Marscher08}, and reduces the number of free parameters in the model. However, we note that the location of particle acceleration has been observed to vary drastically during BHB outbursts \citep{Russell14}; thus $z_{\rm acc} = z_{\rm diss}$ need not be the only viable choice.

Beyond the dissipation region the jet then extends up to a maximum length $z_{\rm max}$. Our code computes the radiation from both the nozzle and extended jet (including synchrotron, synchrotron self-Compton) to reproduce the broadband SED. The synchrotron calculation includes the full individual particle synchrotron spectrum, and the IC calculation accounts for multiple scatterings and the Klein-Nishina cross section. Inverse Compton scattering with other external photon fields such as a torus or broad line region is neglected, as these components are believed to be absent or very faint in low power AGN. We address this choice in section 4.  

If necessary we also include the contribution from the accretion disk to fit the SED, which we model phenomenologically as an optically thick, geometrically thin inflow \citep{Shakura73}, described by an accretion rate $\dot{M}$ normalized in Eddington units, an inner truncation radius $R_{\rm in}$ and an outer radius $R_{\rm out} = 10^{3}\,\rm{R_g}$ (the exact value has a negligible impact on the SED). If a disk contribution is necessary, its photons are included in the Compton calculation as seed photons, but in the case of PKS\,2155$-$304 we find that they are negligible compared to the synchrotron photon field. We assume that emission from the inner disk regions $\left(r<R_{\rm in}\right)$, assumed to be geometrically thick and optically thin, is negligible.  
 
\section{Modelling of PKS\,2155$-$304}
For the first application of our new model we have chosen PKS\,2155$-$304, which is a relatively nearby (redshift $z=0.116$) high-peaked BL Lac. It has been extensively studied by several multi-wavelength campaigns (e.g. \citealt{Hess12}, \citealt{Madejski16}, \citealt{Krauss16}, henceforth K16) that found the source in a variety of spectral states; the wealth of data available makes it an ideal source to benchmark any AGN jet model.

\begin{table}
\centering
\caption{A list of the input parameters of the \texttt{bljet} model. Bold parameters are left free while fitting, parameters marked with an asterisk are only used for some SEDs. Other non-fitted physical parameters are reported after the double horizontal line.}
\begin{tabular}{@{}cp{5.5cm}}
\hline
\hline
Parameter & Description \\
\hline
$\boldsymbol{N_{\rm j}}$ & The total power channelled into the jet base normalized in Eddington units\\
\hline
$\boldsymbol{r_0}$ & The initial radius and aspect ration of the jet nozzle/corona\\
\hline
$\boldsymbol{z_{\rm diss}}$ & The location of the dissipation zone where particle acceleration starts and the jet stops accelerating\\
\hline
$\boldsymbol{\sigma_{\rm diss}}$ & The magnetization at the dissipation region $z_{\rm diss}$, after the jet stops accelerating \\
\hline
$\boldsymbol{p}$ & The slope of the power-law index of accelerated non-thermal particles\\
\hline
$\boldsymbol{f_{\rm heat}}$ & The amount of heating received by the leptons at the start of the dissipation region, which sets the $\gamma_{\rm min}$ of the non-thermal power-law\\
\hline 
$\boldsymbol{f_{\rm b}}$ & A free parameter responsible for setting the dynamical time scale, which fixes the cooling break energy in the lepton distribution\\
\hline
$\boldsymbol{f_{sc}}$ & The efficiency of the particle acceleration process, which sets the maximum energy in the lepton distribution\\
\hline
$\boldsymbol{\dot{M}_{\rm disc}^{*}}$ & The disc accretion rate (when required by data)\\
\hline
$\boldsymbol{R_{\rm in}^{*}}$ & The inner radius of the accretion disc (when required by data)\\
\hline
\hline 
$M_{\rm bh} = 10^9 M_{\odot}$ & The mass of the black hole\\
\hline
$\theta = 2\fdg5$ & The viewing angle between the jet and the line of sight\\
\hline
$h=2r_0$ & The aspect ratio of the jet nozzle/corona\\
\hline
$\gamma_0 = 1.09$ & The initial bulk Lorentz factor of the jet \\
\hline
$\gamma_{\rm acc} = 15$ & The final bulk Lorentz factor of the jet \\
\hline
$\gamma_{\rm th} = 3$ & The peak of the relativistic Maxwellian distribution of thermal leptons\\
\hline 
$\epsilon_{\rm pl} = 0.1$ & The number fraction of leptons accelerated into a non-thermal tail at the dissipation region\\
\hline
$Z_{\rm max}=6.6\cdot10^5\,\rm{R_{g}}$ & The total length of the compact radio jet  where emission is calculated\\
\hline
\hline
\end{tabular}
\label{parlist}
\end{table}

We will be focusing on the datasets produced during extensive monitoring by the TANAMI multi-wavelength program, which involved a variety of instruments operating in different bands (K16). Radio Very Large Baseline Interferometry (VLBI) coverage is provided by the Australian Long Baseline Array (LBA), plus stations in Antarctica, South America and South Africa; in addition, lower resolution observations are performed with the Australian Telescope Compact Array (ATCA) and Ceduna single-dish telescope. These pointings are complemented by data taken with \textit{Swift}/UVOT, the Rapid Eye Mount (REM) telescopes and the Small and Medium Research Telescope System (SMARTS) in the NIR/optical/UV band, \textit{Swift}/XRT in X-rays, and \textit{Fermi}/LAT in high energy (HE) $\gamma$-rays. The details of the data reduction process for each instrument and the production of each SED can be found in K16. Briefly, the \textit{Fermi}/LAT light curve is analysed through a Bayesian block analysis method, with the goal of isolating periods of relatively constant $\gamma$-ray flux, indicating limited variability in each time block. Once these are isolated, an SED is produced by including available pointings of other instruments in each of these periods, which are typically a few months long. The resulting SEDs are quasi-simultaneous: while some variability is expected on much shorter time scales than those probed by the campaign, the overall behaviour of the source is not expected to change dramatically. For PKS\,2155$-$304, the Bayesian block analysis produced six different well sampled SEDs, which following K16 we label $\alpha$, $\beta$, $\gamma$, $\delta$, $\epsilon$, $\zeta$. As shown by the Fermi light curve of K16, reproduced in figure \ref{fermilc}, all of these periods corresponds to low or intermediate states; the only flare detected by \textit{Fermi}/LAT lacks simultaneous pointings of other instruments involved. Because our model represents a steady-state jet, we do not attempt to model short-term variability or flaring states. Instead, our goal is to investigate which of our model parameters are responsible for the long-term variability of the source. The six TANAMI SEDs are ideal for this purpose. 

We assume a systematic error of 10\% for all optical data points (rather than the 2-5 \% reported in K16) because a) the pointings of the three telescopes involved are not strictly simultaneous and b) this reduces the statistical weight of the optical data, compared to radio, X-ray and $\gamma$-ray data, thus leading to a better overall description of the data across all wavelengths. 

In the radio band we only fit the VLBI data and exclude the single dish and ATCA pointings because these low resolution images are likely to be contaminated by the parsec scale jet (resolved out at the VLBI scale) and possibly the radio lobes. 

\subsection{Fitting method}

\texttt{Bljet} is more complex than a one-zone model, and this added complexity introduces degeneracies in our parameter space. Because of this we do not limit ourselves to the typical ``fit-by-eye'' approach used for modelling of blazars. Instead, we perform least-$\chi^{2}$ fits using the Interactive Spectral Interpretation System (ISIS) software package \citep{Houck00}, which allows users to import custom-written models and use them to perform multi-wavelength spectral fitting. Every model is folded through an instrument's response function when available (in our case, this is true for \textit{Swift} data), allowing for a model independent and more precise evaluation of residuals. Like with most $\gamma$-ray satellites, the \textit{Fermi}//LAT PSF is very extended as well as energy-dependent; therefore, sources commonly overlap or contaminate each other; the correct way to treat such data is to do log-likelihood fitting of individual photons. Such behaviour can not be easily treated with tools that have originally been designed for X-ray analysis like ISIS or xspec. This makes including an accurate response function impossible. The main benefit of using ISIS with \textit{Fermi} data is that the software automatically integrates the model flux in each of the (very large) \textit{Fermi} bands, which allows for more accurate comparisons with the data.

We also include an absorption model (\texttt{tbabs}, \citealt{Wilms00}) and a reddening model (\texttt{redden}); for both we fix the column density to the galactic value of $1.48\cdot 10^{20}$ $\rm cm^{-2}$. We adopt the abudances of \cite{Wilms00} and set the photo-ionisation cross-sections according to \cite{Verner96}. The final syntax of the model is: \texttt{tbabs$\times$redden$\times$bljet}. We initially fit the X-ray spectra alone with a power-law plus absorption model; we find that any amount of absorption above the Galactic value is essentially unconstrained by the data.
 
In five out of six epochs we find an excess in the optical bands that cannot be reproduced by our jet model. In order to reproduce it we include a contribution from an accretion disc.

We minimize the residuals with the \texttt{subplex} least-$\chi^{2}$ fitting algorithm included in ISIS. The full list of parameters is provided in Table \ref{parlist}; several of them are frozen either because of observational constraints, or because they do not impact the SED.

The black hole mass in PKS\,2155$-$304 is not measured directly, and estimates based on the host galaxy luminosity range between $1-2\times 10^9 M_{\odot}$ (Aharonian et al. 2007), although when accounting for scatter in the $L_{\rm host} - M_{\rm bh}$ correlation this could be as low as $2\cdot 10^8 M_{\odot}$. We assume a black hole mass of $10^9 M_{\odot}$ and neglect any contribution from the host galaxy to the optical flux. 

We fix the final Lorentz factor of the jet to $15$ and the viewing angle $\theta$ to $2\fdg 5$, which results in a peak Doppler boosting factor of $\delta \approx 22$. In preliminary fits we found that leaving $\gamma_{\rm acc}$ and/or $\theta$ free to vary did not improve the quality of the fits significantly, with the best fit values clustering around these values. Lower terminal velocities and/or larger viewing angles result in very low beaming, making the model incapable of matching the observed fluxes, and in the synchrotron and Compton peaks being shifted to lower frequencies than those observed. Vice versa, for a faster or more beamed jet ($\delta_\gtrsim 30$) the peaks are shifted to higher frequencies, which results in a very poor fit of the X-ray spectra.  

We freeze the maximum length of the compact jet to $10^{20}$ cm, which corresponds to $6.6\cdot 10^{5}\,\rm{R_g}$ for a $10^9 M_{\odot}$ black hole. With this choice, the self-absorption turnover of the outermost region is at $\approx$100 MHz and the spectrum in the GHz frequency range is optically thick and flat (but this is not to suggest the physical jet necessarily ends at this distance).

We assume that $\epsilon_{\rm pl}$, the percentage of particles accelerated into a power-law tail at the dissipation region, is always 10\%. This is consistent with the efficiency expected both for magnetic reconnection and diffusive shock acceleration (\citealt{Sironi11} and \citeyear{Sironi14}, \citealt{Sironi13}, \citealt{Sironi15}). 

Because of relativistic beaming the bulk of the emission originates in regions at $z\geq z_{\rm diss}$, leaving the nozzle mostly unconstrained. Because of this we always take $\gamma_{\rm th} = 3$, which ensures that the thermal synchrotron (and SSC) emission in the nozzle remains negligible. In section 4 we discuss this choice and show that a higher temperature at the base would result in unreasonable features appearing in the SED. The initial radius $r_0$ and nozzle height  $h$ do affect the SED by setting the initial conditions (number density, magnetic field, $z_{\rm 0}$) at the base of the jet. However, the parameter $h$ in particular cannot be constrained without data capable of probing the inner regions of the inflow/outflow (such as a reflection signature in the X-ray spectra), and therefore we assume $h=2r_0$ and leave the initial radius $r_0$ as a free parameter.

These choices leave us with 8-10 free parameters (depending on whether we include a contribution from the disc), described in table \ref{parlist}.

\begin{table*}
\caption{Best fit parameters for every SED fit individually. Fitted free parameters are bolded; we also report the values calculated by the model for the magnetic field, non-thermal lepton number density (only the non-thermal particles contribute meaningfully to the SED in this source), minimum, break and maximum energies reached by the leptons, all computed at a distance of $3\cdot 10^3$\,$\rm R_{\rm g}$ from the black hole, as well as the reduced $\chi^{2}$ for each fit. These parameters require an initial magnetization $\sigma_{\rm 0}\approx 13$. We do not report any confidence intervals for the individual fits because the parameter space of individual fits is too degenerate (see Fig.\,\ref{singlemcmc}).}
\begin{tabular}{| l | c | c | c | c | c | c | c  | c | c | c | c | c | c | c | c | c | r |}
\hline
\hline
& $\boldsymbol{\dot{M}_{\rm disc}}$ & $\boldsymbol{R_{\rm in}}$ & $\boldsymbol{N_{\rm j}}$ & $\boldsymbol{r_0}$  & $\boldsymbol{z_{\rm diss}}$ & $\boldsymbol{p}$ & $\boldsymbol{f_{\rm heat}}$ & $\boldsymbol{f_{\rm b}}$ & $\boldsymbol{f_{\rm sc}}$ & $\boldsymbol{\sigma_{\rm diss}}$ &  $B$ & $n(e)$ & $\gamma_{\rm min}$ &  $\gamma_{\rm brk}$  & $\gamma_{\rm max}$ & $\chi^{2}/{\rm dof}$ \\
& [$\rm \dot{M}_{\rm Edd}$] & [$\rm R_{g}$] & [$\rm L_{\rm Edd}$] &  [$\rm R_{g}$] & [$\rm R_{g}$] &  &  &  & $10^{-6}$ & $10^{-2}$ & [$\rm G$] & [$\rm cm^{-3}$] & & $10^{2}$ & $10^{5}$ &\\
& $10^{-2}$ &  & $10^{-2}$&  &  &  &  &  & & & & & & & &\\
\hline
$\alpha$ & 2.9 & 22 & 1.0 & 29 & 510 & 1.9 & 20 & 40 & 2.7 & 2.0 & 0.24 & 52.6 & 133 & 11 & 3.2 & 59.33/22\\
\hline
$\beta$ & 1.6 & 30 & 1.0 & 75 & 1360 & 1.8 & 10 & 22 & 2.5 & 3.2 & 0.30 & 17.3 & 63 & 7.1 & 2.7 & 38.51/15\\
\hline
$\gamma$ & / & / & 0.9 & 23 & 1700 & 1.9 & 11 & 30 & 2.0 & 5.6 & 0.37 & 56.6 & 72 & 7.1 & 2.4 & 58.3/21\\
\hline
$\delta$ & 1.4 & 100 & 0.9 & 10 & 1170 & 1.6 & 8 & 86 & 1.2 & 2.0 & 0.23 & 97.3 & 51 & 7.3 & 1.8 & 66.05/24\\
\hline
$\epsilon$ & 0.7 & 79 & 1.5 & 15 & 960 & 1.8 & 6 & 50 & 1.2 & 1.4 & 0.23 & 130 & 43 & 8.2 & 1.8 & 29.87/24\\
\hline
$\zeta$ & 0.9 & 18 & 1.6 & 26 & 1720 & 1.9 & 8 & 43 & 1.6 & 1.6 & 0.28 & 90.1 & 55 & 8.1 & 2.3 & 33.72/25\\
\hline
\hline
\end{tabular}
\label{singlefitpars}
\end{table*}

\begin{figure*}
\hspace*{-0.5cm}
\includegraphics[scale=.7]{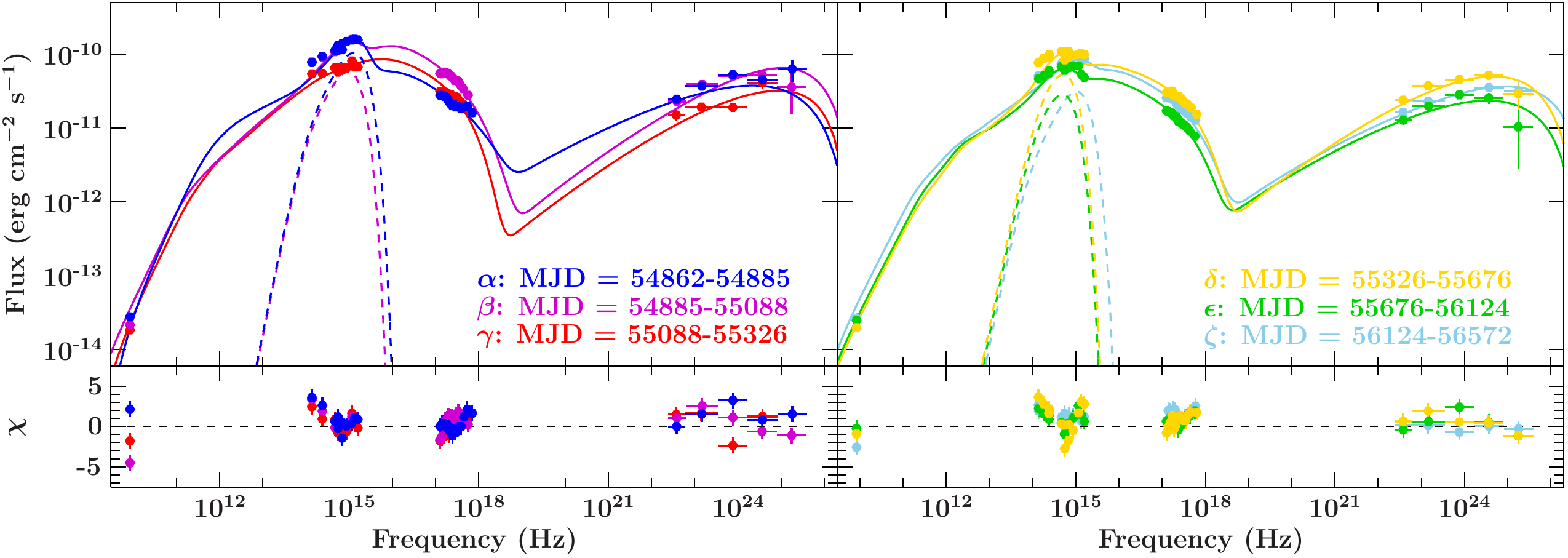}
\caption{
Individual fits to the six SEDs in the top two panels, with residuals shown in the bottom two. The contribution from the accretion disc at each epoch is shown by the dashed lines.
}
\label{single}
\end{figure*}

\begin{figure*}
\hspace*{-0.5cm}
\includegraphics[height=0.2\textwidth]{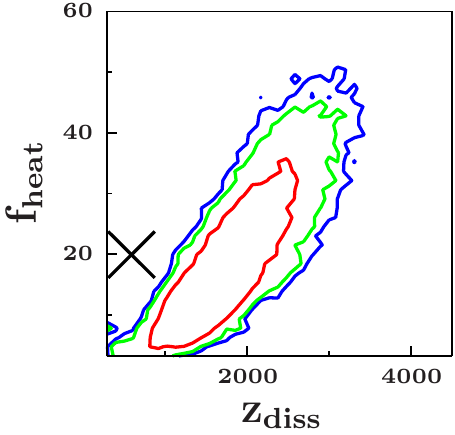}
\includegraphics[height=0.2\textwidth]{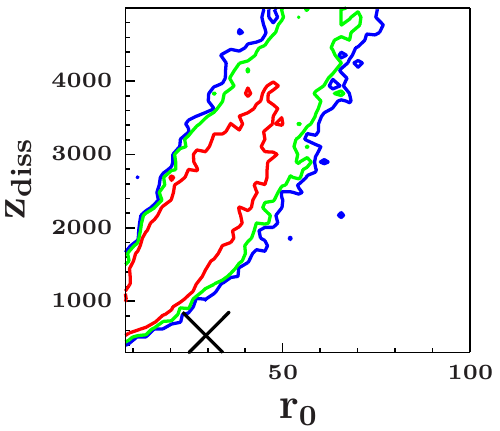}
\includegraphics[height=0.2\textwidth]{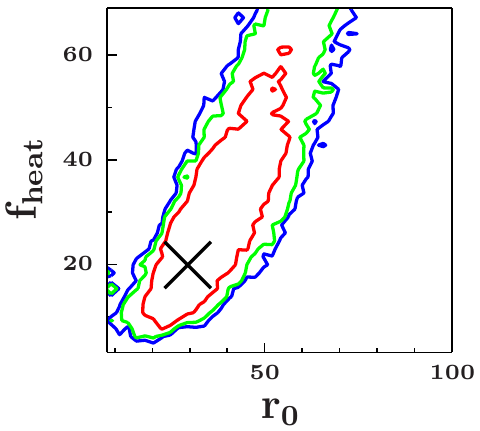}
\includegraphics[height=0.2\textwidth]{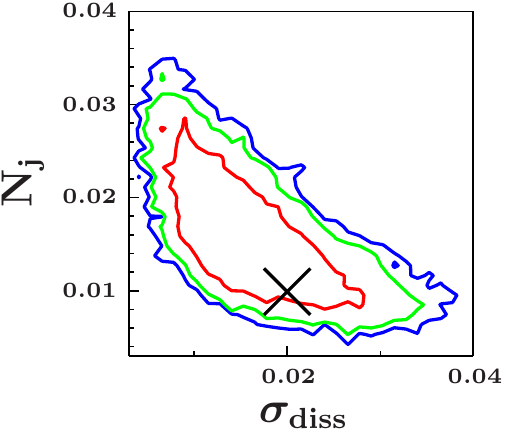}
\caption{
Degeneracy from individual fits: the plots show the most two-dimensional contour plots for the $\alpha$ SED which show a correlation between parameters. The red, green and blue contours indicate 68, 90 and 95 per cent confidence intervals found by \texttt{emcee}; the black cross denotes the best fit values found by the least $\chi^{2}$ routine. We find strong correlations between the initial radius $r_{\rm 0}$ and location of the dissipation region $z_{\rm diss}$, injected jet power $N_{\rm j}$ and final magnetization $\sigma_{\rm diss}$, particle heating $f_{\rm heat}$ and $z_{\rm diss}$, all of which are very poorly constrained. 
}
\label{singlemcmc}
\end{figure*}

\subsection{Individual fits}

Our best fits to individual datasets are shown in Fig.~\ref{single}, and the best fit parameters are reported in Table \ref{singlefitpars}, along with the values of the magnetic field, lepton density, minimum and maximum Lorentz factors reached by the emitting particles at $z = 3 \cdot 10^3$\,${\rm R}_{\rm g}$ (the regions around this distance are responsible for a large contribution to the SED, with the exception of the radio emission), and $\chi^{2}/{\rm dof}$ of the best fit parameters. In all cases, the broadband SED is described very well by the model.

In every dataset except $\gamma$ an additional thermal component is required to match the optical flux and spectral shape. In the case of $\alpha$ and $\epsilon$ the excess optical bump is easily seen in the data. A less visible bump is also present in $\beta$, $\delta$ and $\zeta$. This additional variable thermal component was also found by K16, who modelled the same datasets with phenomenological two log-parabolas (plus a black body if necessary). We model this additional component as a contribution from a truncated geometrically thin, optically thick accretion disk, neglecting any emission from regions at radii smaller than the truncation radius. We find that both the accretion rate and truncation radius have to vary between epochs in order to match the optical excess. We address the inferred disk variability in section 4.

The most notable trend emerging from these individual fits is that the bulk acceleration process always lasts until the magnetization $\sigma$ is smaller than 1 and of the order of $\approx 2-3 \cdot 10^{-2}$, meaning that the jet always transitions from a Poynting-dominated base ($\sigma \gg 1$) to a kinetic-dominated outer region ($\sigma \ll 1$) as it accelerates. This transition is also consistently required when modelling the SEDs of TeV BL Lacs with one-zone models (e.g. \citealt{Tavecchio16}). Despite this departure from equipartition between kinetic and magnetic powers, we always find that the required jet power is sub-Eddington, and comparable to the accretion rate inferred from modelling the optical data. This finding is consistent with those of \cite{Ghisellini14}, who found a strong correlation between accretion rate (inferred from the luminosity of the BLR) and jet power (measured through SED fitting with a one-zone model), with the latter being of the same order of magnitude but systematically higher. We note however that their study is limited to high power blazars in which the BLR is clearly detected, which is not the case for PKS\,2155$-$304.

In three out of six epochs ($\gamma$, $\epsilon$, $\zeta$) the power injected at the base of the jet is higher than the inferred accretion rate in the outer thin disk. As we will show in the next section this is purely a result of the degeneracy of the model. We note that in most cases the contribution of the disc is neglected in PKS\,2155$-$304 due to its featureless optical spectrum (e.g. \citealt{Hess12}), which means that the true accretion rate of this source is presently unknown. However, if the accretion rate is of the order of the estimated jet power ($\approx 10^{-2}$ $L_{\rm Edd}$) then the disc could possibly contribute in some amount to the observed SED, as required by our model.

Finally, we note that the main parameter driving changes in the (optically thick) radio flux predicted by the model is the injected jet power $N_{\rm j}$. A multi-zone model capable of fitting this part of the spectrum should therefore constrain the jet power more effectively that a one-zone model, as long as model degeneracy is limited or accounted for. Our values for the jet power are consistent with the lower limit estimated from \textit{NuSTAR} observations by \cite{Madejski16}. 

\subsection{Model degeneracies and joint fitting}

After fitting the six datasets using $\chi^{2}$ minimization we further explore our parameter space using \texttt{emcee}, an ISIS implementation \citep{Murphy14} of the Monte Carlo Markov Chain (MCMC) method of \cite{Foreman13}. \texttt{Emcee} sets up a distribution of ``walkers'' which then explore the parameter landscape: for each iteration the walkers jumps to a new parameter value, and depending on the $\chi^{2}$ values in the new and old position the move may be accepted or rejected. For each \texttt{emcee} run we initialize 100 walkers per free parameter in a narrow Gaussian distribution around the best fit values found, where the standard deviation is 1\% of the best fit value. We choose a Gaussian rather than flat distribution because in trial runs we found this results in faster convergence of the chain. The output of \texttt{emcee} allows us to identify possible modelling degeneracies which could force the least-$\chi^{2}$ in a local rather than global minimum, and also to estimate error bars for the best-fit parameters. We found that for an individual SED the chain takes around one thousand iterations to converge to a good fit, but it identifies possible correlations between parameters in a few hundred steps.

We initially run \texttt{emcee} for the $\alpha$ dataset exclusively with the goal of identifying degeneracies in our model, and therefore only evolve the chain for 1000 iterations. We take the first 200 iterations as a ``burn-in'' period and discard them. Fig.~\ref{singlemcmc} shows the four main correlations found among the 10 free parameters in the final 800 steps. We find that the parameters that show significant correlations are: the jet power $N_{\rm j}$ and final magnetization $\sigma_{\rm diss}$; the initial radius $r_0$, dissipation distance $z_{\rm diss}$, and electron heating $f_{\rm heat}$; and finally the accretion rate $\dot{M}$ and inner disc radius $R_{\rm in}$ (not shown). Due to these degeneracies the chain does not recover the same best fit values as the least-$\chi^{2}$ method, and at the same time the quality of the fit does not improve significantly, implying that the parameter space is too complex and multi-modal to estimate parameter uncertainties; however, the least-$\chi^{2}$ and MCMC fits are roughly consistent with each other. 

Our method to attempt to break these degeneracies is to perform \textit{joint} SED fitting: the datasets are loaded simultaneously and a separate instance of the chosen spectral model is assigned to each dataset, with several parameters tied across every dataset rather than being left to vary independently. This approach has been used successfully to study individual SEDs of LLAGN and BHBs simultaneously (\citealt{Markoff15}, \citealt{Connors17}), but it has never been applied to different multi-wavelength datasets of the same source until now (but see \citealt{Connors18}, submitted, for a similar study of the BHB GX~339$-$4).

\begin{table*}
\caption{Best joint fit parameters (bolded). We also include the calculated values for break and maximum energies reached by the accelerated leptons at a distance of $3\cdot 10^3$ $\rm R_{\rm g}$ from the black hole. At this distance we find a magnetic field of 0.25 G, a non-thermal lepton number density of 70 $\rm cm^{-3}$ and a minimum Lorentz factor of 69; as in the individual fits the initial magnetization is $\sigma_0\approx 13$. We also report the reduced $\chi^{2}$ for the full joint fit. Unlike the individual fits, the parameter space is relatively well-behaved, and thus we can report confidence intervals along for the best-fit parameters.}
\hspace*{-0.75cm}
\begin{tabular}{| l | c | c | c | c | c | c | c | c | c | c | c | c | c | c | c | c | r |}
\hline
\hline
& $\boldsymbol{\dot{M}_{\rm disc}}$ & $\boldsymbol{R_{\rm in}}$ & $\boldsymbol{N_{\rm j}}$ & $\boldsymbol{r_0}$  & $\boldsymbol{z_{\rm diss}}$ & $\boldsymbol{p}$ & $\boldsymbol{f_{\rm heat}}$ & $\boldsymbol{f_{\rm b}}$ & $\boldsymbol{f_{\rm sc}}$ & $\boldsymbol{\sigma_{\rm diss}}$ &  $\gamma_{\rm brk}$  & $\gamma_{\rm max}$ & $\chi^{2}/{\rm dof}$ \\
& [$\rm \dot{M}_{\rm Edd}$] & [$\rm R_{g}$] & [$\rm L_{\rm Edd}$] &  [$\rm R_{g}$] & [$\rm R_{g}$] &  &  &  & $10^{-6}$ & $10^{-2}$ & $10^{2}$ & $10^{5}$ &\\
& $10^{-2}$ & & $10^{-2}$ & & &  &  &  & & & & &\\
\hline
Joint &  &  & $0.90^{+0.06}_{-0.07}$ & $18^{+3}_{-2}$ & $600^{+62}_{-65}$ &  & $10.4^{+0.8}_{-0.6}$ &  &  & $2.5^{+0.1}_{-0.2}$ &  &  & 265.54/156\\
\hline
$\alpha$ & $2.6^{+0.3}_{-0.2}$ & $20^{+3}_{-2}$ &  &  &  & $1.74^{+0.05}_{-0.04}$ &  & $48^{+12}_{-4}$ & $1.4^{+0.1}_{-0.2}$ &  & 8& 2.0 & \\
\hline
$\beta$ & $2.6^{+0.3}_{-0.3}$ & $46^{+8}_{-7}$ &  &  &  & $2.01^{+0.04}_{-0.03}$ &  & $8^{+2}_{-1}$ & $4.2^{+0.8}_{-0.7}$ &  & 52 & 3.3 & \\
\hline
$\gamma$ & $0.8^{+0.1}_{-0.1}$ & $23^{+4}_{-3}$ &  &  &  & $1.99^{+0.04}_{-0.03}$ &  & $17^{+3}_{-3}$ & $5.1^{+0.6}_{-0.4}$ &  & 24 & 3.8 & \\
\hline
$\delta$& $1.4^{+0.1}_{-0.1}$ & $110^{+20}_{-20}$ &  &  &  & $1.90^{+0.03}_{-0.03}$ &  & $17^{+3}_{-1}$ & $1.9^{+0.1}_{-0.1}$ &  & 23 & 2.3 & \\
\hline
$\epsilon$ & $0.8^{+0.1}_{-0.1}$ & $79^{+10}_{-11}$ &  &  &  & $1.98^{+0.03}_{-0.03}$ &  & $17^{+3}_{-2}$ & $1.3^{+0.1}_{-0.1}$ &  & 22 & 1.9 & \\
\hline
$\zeta$ & $1.2^{+0.1}_{-0.1}$ & $31^{+5}_{-4}$ &  &  &  & $1.94^{+0.03}_{-0.03}$ &  & $20^{+4}_{-3}$ & $2.2^{+0.2}_{-0.1}$ &  & 20 & 2.5 & \\
\hline
\hline
\end{tabular}
\label{jointfitpars}
\end{table*}

\begin{figure*}
\hspace*{-0.5cm}
\includegraphics[scale=.7]{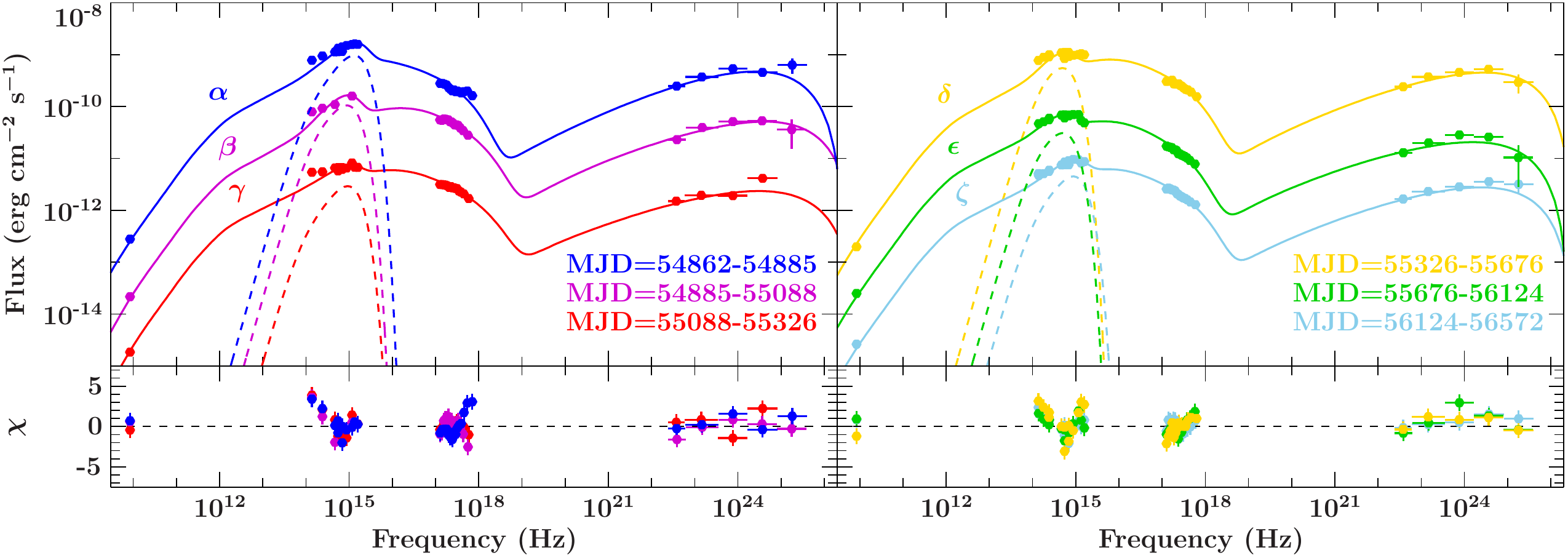}
\caption{
Joint fits to the six SEDs in the top two panels, with residuals shown in the bottom two. Four SEDs have been shifted up ($\alpha$, $\delta$) or down ($\gamma$, $\zeta$) by a factor of 10 for clarity. Despite the stronger constraints imposed by jointly fitting all the SEDs, the model remains in excellent agreement with the data. 
}
\label{joint}
\end{figure*}

We choose to tie all the jet parameters which show degeneracy: $N_{\rm j}$, $r_0$, $z_{\rm diss}$, $\sigma_{\rm diss}$ and $f_{\rm heat}$. Physically, this corresponds to assuming that the bulk source properties (jet dynamics, shape and energy budget) are unchanged over the time scales probed by the our data. The time scale over which we might expect these properties to vary is roughly $t_{\rm var} \approx z_{\rm max}/\left[c\delta\left(1+z\right)\right]\approx 1.5\,\rm{yr}$, comparable to the TANAMI sampling. This suggests that even if the bulk properties of the source did change with time their variation should be relatively small, justifying our assumption to tie them across epochs. Unlike the bulk jet properties, we cannot address the degeneracy between the two disc parameters, as we find they need to be different in different epochs (and entirely absent in one), which prevents us from tying them. We discuss the implications of the inferred disk variability in the following section. Fig.~\ref{joint} shows the result of our best joint fit for all datasets; the best fit parameters are reported in Table \ref{jointfitpars}. The total number of fully free parameters in each SED is now 5, plus the 5 that have been tied. We find that despite the additional constraint imposed by the joint fit, the model remains in excellent agreement with the data at all epochs, as shown in Fig.~\ref{joint}. The X-ray slope of the $\alpha$ dataset and the NIR/optical slopes at all epochs show slightly worse than individual fits (with the structure seen in the residuals being similar in both individual and joint fits), but the data are still well reproduced. We also point out that the $\alpha$ state is both the lowest X-ray state and brightest $\gamma$ ray state identified during the TANAMI monitoring of the source (see also the SED plots in K16), making it the most constraining (and challenging) dataset to model. As a consistency check we also tried running one more fit in which the tied parameters were untied again and allowed to vary within 10\% of the value found during the joint fit, but this did not improve the quality of the fit. Finally, because our model includes only a phenomenological treatment of both the non-thermal particle distribution and the accretion disk, we do not consider this discrepancy between the model and the data to invalidate the joint fit found.

The 5 degenerate parameters fall within the range allowed by the individual fits, showing that the joint fit recovers the same physics but also discriminates more effectively  between the various degenerate solutions allowed by the individual fits. This is the key result of this study. In particular, we find that the differences in the SED at different epochs can be reproduced mainly by varying the slope, break energy and maximum energy of the radiating particle distribution, while the bulk properties of outflow remain unchanged.

\begin{figure*}
\hspace*{-0.5cm}
\includegraphics[height=0.2\textwidth]{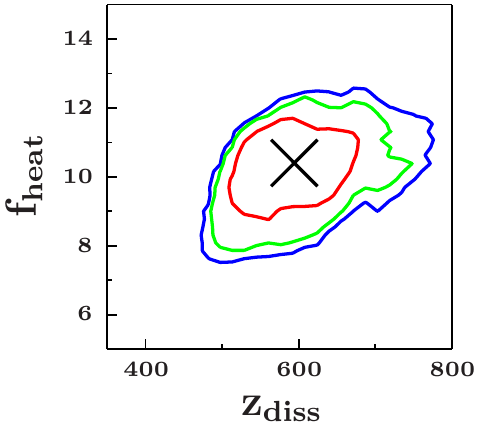}
\includegraphics[height=0.2\textwidth]{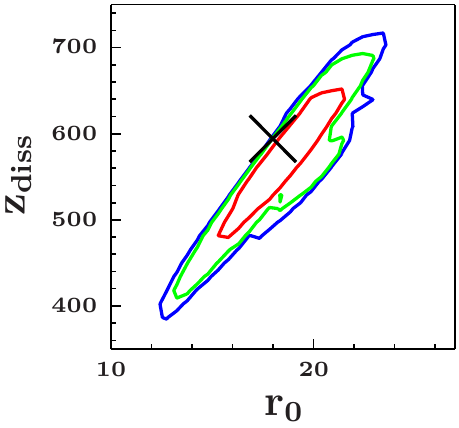}
\includegraphics[height=0.2\textwidth]{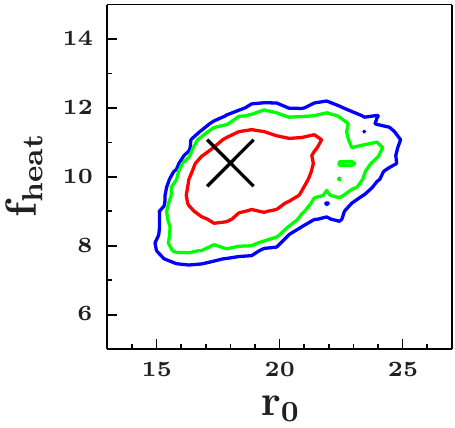}
\includegraphics[height=0.2\textwidth]{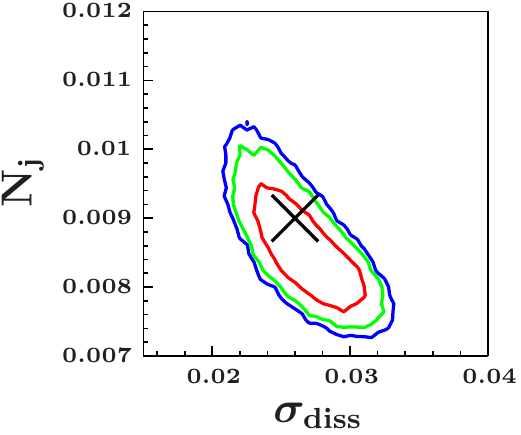}
\caption{
Two-dimensional confidence contour histograms for the degenerate parameters in our model for the joint fit. The red, green and blue contours indicate 68, 90 and 95 per cent confidence intervals found by \texttt{emcee}; the black cross denotes the best fit values found by the least $\chi^{2}$ routine. Unlike in the individual fit, the degeneracy between $f_{\rm heat}$ and the other parameters almost completely disappears, and every tied parameter is well constrained.
}
\label{jointmcmc}
\end{figure*}

We run a final \texttt{emcee} routine for the full joint fit in order to identify any remaining degeneracies. A full exploration of parameter space for all six SEDs, each with its own instance of the model, is extremely computationally intensive, and therefore we cannot evolve this chain for more than 1000 loops\footnote{The final chain took about 6 weeks on a 32-core AMD Ryzen Threadripper 1950X CPU, using 30 slave processes.}. As in the first chain, we initialize 100 walkers per free parameter in a narrow Gaussian distribution around the best-fit values found during least-$\chi^{2}$ fitting. We discard the first 200 loops as the ``burn-in'' period. The final contour plots for the tied parameters are shown in Fig.~\ref{jointmcmc}. 

\begin{figure}
\begin{center}
\includegraphics[scale=0.7]{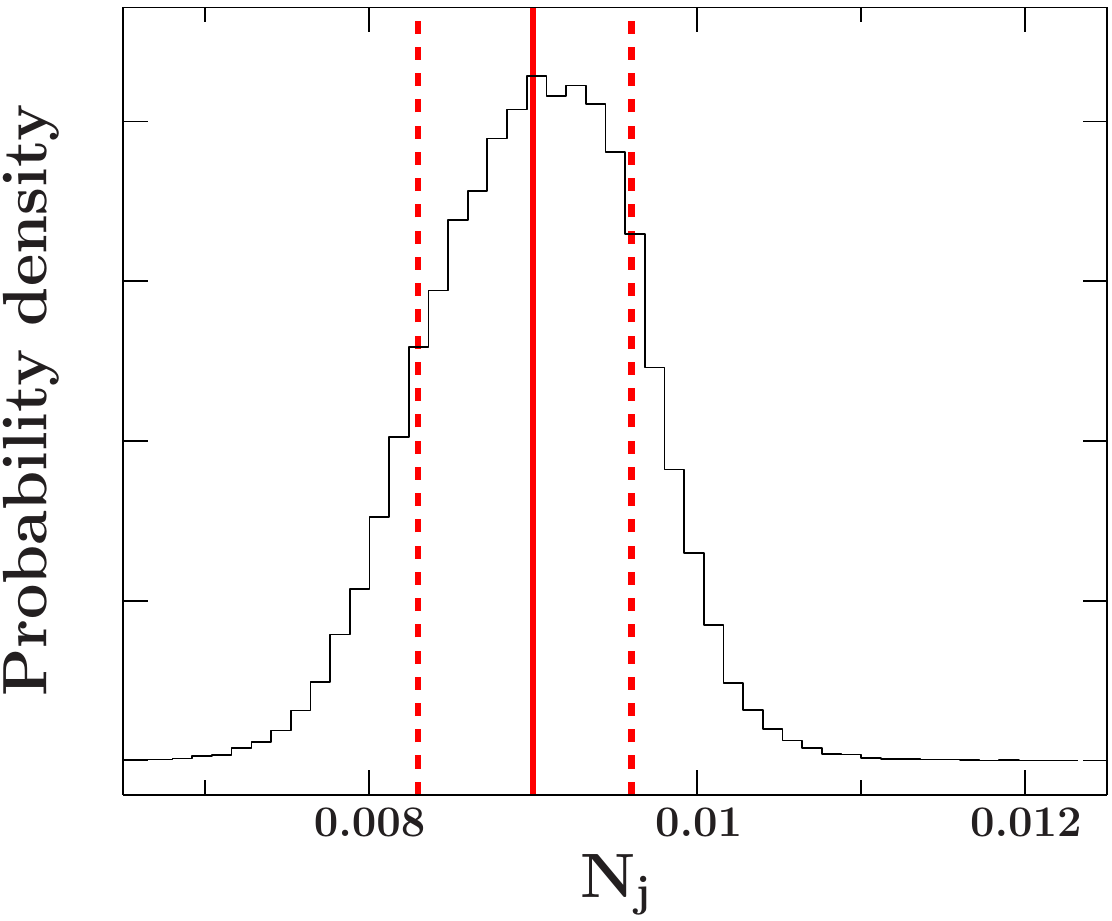}
\end{center}
\caption{
Posterior distribution for the injected jet power $N_{\rm j}$ found during the joint \texttt{emcee} run. The red continuous line denotes the maximum of the posterior distribution, the dashed red lines the 68\% confidence intervals.
}
\label{njdist}
\end{figure}

The new \texttt{emcee} chain confirms that the joint fit is far more constraining than the individual fits. We find that, unlike in the single-SED run, 1000 loops are enough for the chain to converge and recover the best-fit parameter values, in the sense that the peak of the posterior distribution found by the \texttt{emcee} and the value found by the least-$\chi^{2}$ algorithm are in agreement with each other. This result demonstrates that the parameter space for the joint fits is far smaller and less multi-modal than for the individual fits. Because the chain successfully converged to a good fit very quickly (around 200 loops) we can use its output to estimate confidence intervals, which we define as the intervals in the the one-dimensional histograms containing 68\% of the walkers from the end of the burn-in period to the end of the \texttt{emcee} run. An example of such a histogram for the jet power $N_{\rm j}$ is shown in Fig.~\ref{njdist}. We also find that the degeneracy of the heating parameter $f_{\rm heat}$ almost completely disappears. This results in the allowed intervals for $r_0$ and $z_{\rm diss}$ being far smaller, to the point where despite the inherent degeneracy between these two parameters they are rather well constrained. A similar behaviour is also observed for the jet power $N_{\rm j}$ and the magnetization at the dissipation region $\sigma_{\rm diss}$: while the two parameters remain degenerate with each other, they are much better constrained.

\section{Discussion}

The main trend emerging from our joint fitting approach using the multi-zone jet model is that the long term variability of the source can be reproduced by changing the details of the non-thermal particle distribution (in particular $p$, $\gamma_{\rm b}$, $\gamma_{\rm max}$) while keeping the bulk jet parameters (geometry, magnetization, injected power) unchanged. This result implies that at least outside of flaring states, the outflow is in a steady-state configuration but the local properties of the plasma are varying, leading to changes in efficiency of the acceleration mechanism responsible for producing the non-thermal radiating particles.

\begin{figure}
\hspace{-1.0cm}
\includegraphics[scale=.35]{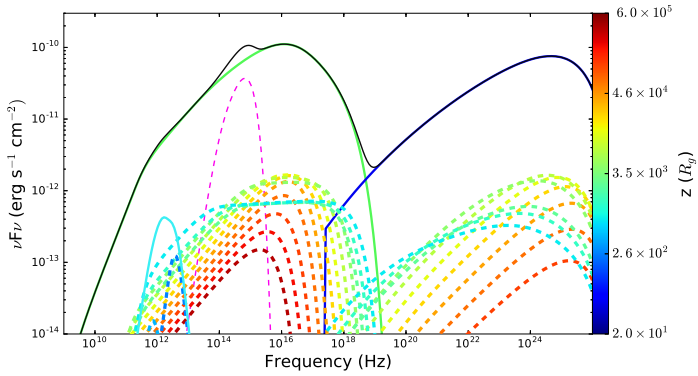}
\caption{
Typical SED computed by our model; the teal/green/blue continuous lines represent the total thermal synchrotron, non-thermal synchrotron, and inverse Compton emission, the dashed magenta line the disc emission, and the black continuous line the integrated flux. The coloured dashed lines show the contribution  at varying distance from the black hole. Light blue lines correspond to inner ($z < 1500\,\rm{R_g}$) regions of the jet, green/yellow lines to intermediate regions ($z \approx 10^{3}-10^{4}\,\rm{R_g}$), and orange/red to outer regions ($z \approx 10^{4}-10^{5}\,\rm{R_g}$).
}
\label{zones}
\end{figure}

Figure \ref{zones} shows a typical SED of the source, as well as the individual contribution from several zones. We find that the particle distribution is very strongly cooled in regions relatively close to the base ($z \leq 2000\,\rm{R_g}$) due to the strong magnetic field present in this region, suppressing the contribution from these zones to the SED. The bulk of the emission originates at intermediate distances ($z \approx 10^{3}-10^{4}\,\rm{R_g}$), where the magnetic field is low enough to not cause strong cooling, and the number density of particles is still relatively high, resulting in fairly bright emission. In particular, most of the SSC emission is originated in this section of the jet. Interestingly, the strong effect of cooling in the inner regions of the jet implies that the bulk of the emission comes from regions farther downstream from where the jet stops accelerating. The outer regions ($z \geq 10^{4}\,\rm{R_g}$) mostly contribute in the radio band; their IC emission is so faint that we neglect its calculation to speed up our code. 

In every SED except one (epoch $\gamma$ in the individual fits), we require an additional component to match the optical emission, which we model as a variable truncated optically thick, geometrically thin accretion disc. In order to match the optical excess we find that the truncation radius $R_{\rm in}$ has to vary between epochs by a factor of $\approx\,6$, and that the best-fit values at each epochs are statistically inconsistent with each other. The truncation radius should vary only over a viscous time scale (\citealt{Done07}, \citealt{Yuan14}, and references therein), which for a disk at a distance of $50\,\rm{R_{g}}$ from a black hole of $10^{9}\,\rm{M_{\odot}}$ is around $10^{3}$ yr \citep{Frank02}. This time scale is far larger than those probed by our data, implying that if a disk contribution is indeed present in the SED a simple truncated disk is too simplistic to fully capture its physics. One possible explanation is that the inferred disk variability is not caused by a variation in the truncation radius, but by disk irradiation from a central X-ray source instead. When the central source varies (which can happen on time scales far smaller than the viscous time scale), so does the reprocessed disk emission -- \cite{Gierlinki08} showed that this process can lead to inferring a large variation on the disk truncation radius, as is the case in our SEDs. Unfortunately the total contribution of the accretion flow to the SED is not sufficient to constrain this scenario. Accurate modelling of the accretion flow in this source is beyond the scope of this paper, so in the following discussion we will assume that the ``true'' parameters of the accretion disk are of the same order of magnitude as those we found while fitting the data. We also note that modelling the source during a single epoch would not have highlighted the potential issue of disk variability.

We find that the jet power and accretion rate are roughly of the same order of magnitude in both individual and joint fits. Individual fits for three epochs ($\gamma$, $\epsilon$, $\zeta$) require higher jet powers than accretion rates, thus implying an additional source of energy needed to power the jet such as the black hole's spin \citep{Blandford77}. The trend of higher jet power with respect to accretion rate however is not seen in the more constraining joint fit, implying that it is exclusively a product of the model's degeneracy. Our model therefore cannot discriminate black hole spin \citep{Blandford77} from disc angular momentum \citep{Blandford82} as the origin of jet powering for PKS\,2155$-$304. We also point out that jet power estimates based on SED fitting are very strongly model-dependent. For example, \cite{Madejski16} find a range between $10^{45}$ and $10^{47}$\,erg\,s$^{-1}$ depending on the jet composition and lepton distribution, while \cite{Potter13b} find $1.6\times 10^{44}$\,erg\,s$^{-1}$ but do not include a proton contribution to the jet energy budget. The main constraint on the jet power in our model ($1.24^{+0.08}_{-0.09}\cdot 10^{45}\,\rm{erg\,s^{-1}}$) is given by the radio flux, which is rarely fit by one-zone models. Our statistical analysis over multiple epochs therefore provides a strong constraint on this quantity and allows for a very small range of values, which we find to be in agreement with the lowest value allowed in \citeauthor{Madejski16}'s \citeyear{Madejski16} work.

The best fit values for our model require the jet to accelerate strongly over a small distance, particularly for the joint fit ($z_{\rm diss} = 600^{+62}_{-65}$\,$R_{\rm g}$). This conclusion is roughly consistent with one-zone models of other blazars, but not consistent with VLBI observations of M87, for which the jet geometry and acceleration profile can be mapped from scales of a few $R_{\rm g}$ up to the parsec- and kilo-parsec- scale. In M87, the transition region from an accelerating parabolic flow to a roughly conical one is seen to occur at around $10^{5}$\,$R_{\rm g}$ (e.g. \citealt{Biretta99}, \citealt{Asada12}, \citealt{Hada13}, \citealt{Hada16}). If such a source were to be seen face-on at a cosmological redshift, according to our model its  emission would be very faint compared to PKS\,2155$-$304 (particularly in the $\gamma$-ray band) unless the jet power and Doppler factor were extremely high, as the non-thermal particles would be injected in a region of very low magnetic fields and particle density (or modest beaming and high magnetization, if particle injection were to occur closer to the base). While the SED of M87 is relatively similar to that of a typical low power blazar (\citealt{Tavecchio08}, \citealt{deJong15}), the dynamics of its jet may not be. A similar trend of jet acceleration lasting up to large distances is also seen in GRMHD simulations (\citealt{McKinney06}, Chatterjee et al., in prep). A possible way to resolve this tension could come from a comparison of jetted AGN and X-ray binaries. During black hole X-ray binary outbursts the synchrotron self-absorption break in the jet is seen to vary by several orders of magnitude between epochs while the source is still in the hard state \citep{Russell14}. Such time-dependent behaviour implies that key properties of the jet, such as the jet acceleration and particle injection regions, can change dramatically over time for the same black hole, at similar accretion rates. If super-massive black holes behave in the same way as stellar mass black holes on longer time scales, then the jets of M87 and PKS\,2155$-$304 could simply be in different ``configurations/states'' despite both being low-power, jetted sources. We also note that recent \textit{RadioAstron} observations of the FR\RN{1} radio galaxy 3C84 \citep{Giovannini18} show a very different jet geometry from M87, further strengthening the suggestion that black hole jets in different sources can have very different dynamics and structure. 

Our findings for the location of the jet dissipation region are in contrast to those of \cite{Potter13b}, who modelled PKS\,2155$-$304 (J2158.8$-$3014 in their work) with a similar multi-zone jet model which includes magnetic acceleration. In their work they assume that the jet geometry is the same as that of M87, with a transition between the accelerating, parabolic inner flow and the conical, slowly decelerating (in their model) region at $10^{5}$\,$R_{\rm g}$; if necessary, they vary the black hole's mass in order to rescale their model. Unlike in our model, they self-consistently account for turning bulk kinetic energy into internal energy through shocks. Instead, our model does not account for the additional energy required to heat and accelerate the electrons. However, because the cold protons dominate the overall particle energy budget, this additional energy is small and our estimate for the injected jet power is close to the true jet power.

For PKS\,2155$-$304 they achieve a good fit by scaling the black hole mass to $2.3 \cdot 10^{7}$\,$M_{\odot}$, one to two orders of magnitude lower than that inferred for the source \citep{Aharonian07}. They find a power of $\approx 10^{-1}$\,$\rm{L_{Edd}}$, which is one order of magnitude higher than in our model, despite their choice of a light, pair-dominated jet. This is because in our model the highly beamed regions dominating the emission are much closer to the black hole, approximately between $600$ and $10^{4}$\,$R_{\rm g}$, where the particle density and magnetic fields are higher, resulting in brighter emission despite the lower initial energy budget. This comparison between our jet model and theirs provides another hint that the structure and dynamics of the jet in M87 are likely different from those of a canonical blazar like PKS\,2155$-$304.

We find that the jet in PKS\,2155$-$304 has to become strongly particle-dominated at the dissipation region($\sigma_{\rm diss} \approx 0.02$, including cold protons) in order to match the \textit{Fermi}/LAT data, which we reproduce purely through SSC. This happens because the bolometric synchrotron luminosity scales as $L_{\rm S} \propto n$, where $n$ is the number density of the radiating particles, while for SSC $L_{\rm SSC} \propto n U_{\rm rad}^{\prime} \propto n^{2}$; therefore, $L_{\rm SSC}/L_{\rm S}\propto n$. In our definition the magnetization $\sigma \propto 1/n$ (Eq.~\ref{sigma}), which implies $L_{\rm SSC}/L_{\rm S}\propto 1/\sigma$: if the Compton bump is due to SSC the magnetization has to be low in order to match the observed flux. Our result is consistent with one-zone models for TeV-detected BL Lacs \citep{Tavecchio16}. We note however that in our model the average Lorentz factor of the radiating leptons is relatively low ($\langle\gamma_e\rangle \approx 10^{2}$), meaning that their energy density is far lower than that of the protons. Low values of $\sigma_{\rm diss}$ therefore mean that in our model for this source $U_{\rm p} \gg U_{\rm b} \approx U_{\rm e}$. As a result, deviation from ``equipartition'' is far less severe than that reported by \citeauthor{Tavecchio16} (\citeyear{Tavecchio16}) with one-zone models, which in their work require $\langle\gamma_e\rangle \geq 10^{3}$. 

\begin{figure*}
\hspace{-1.cm}
\includegraphics[scale=.3]{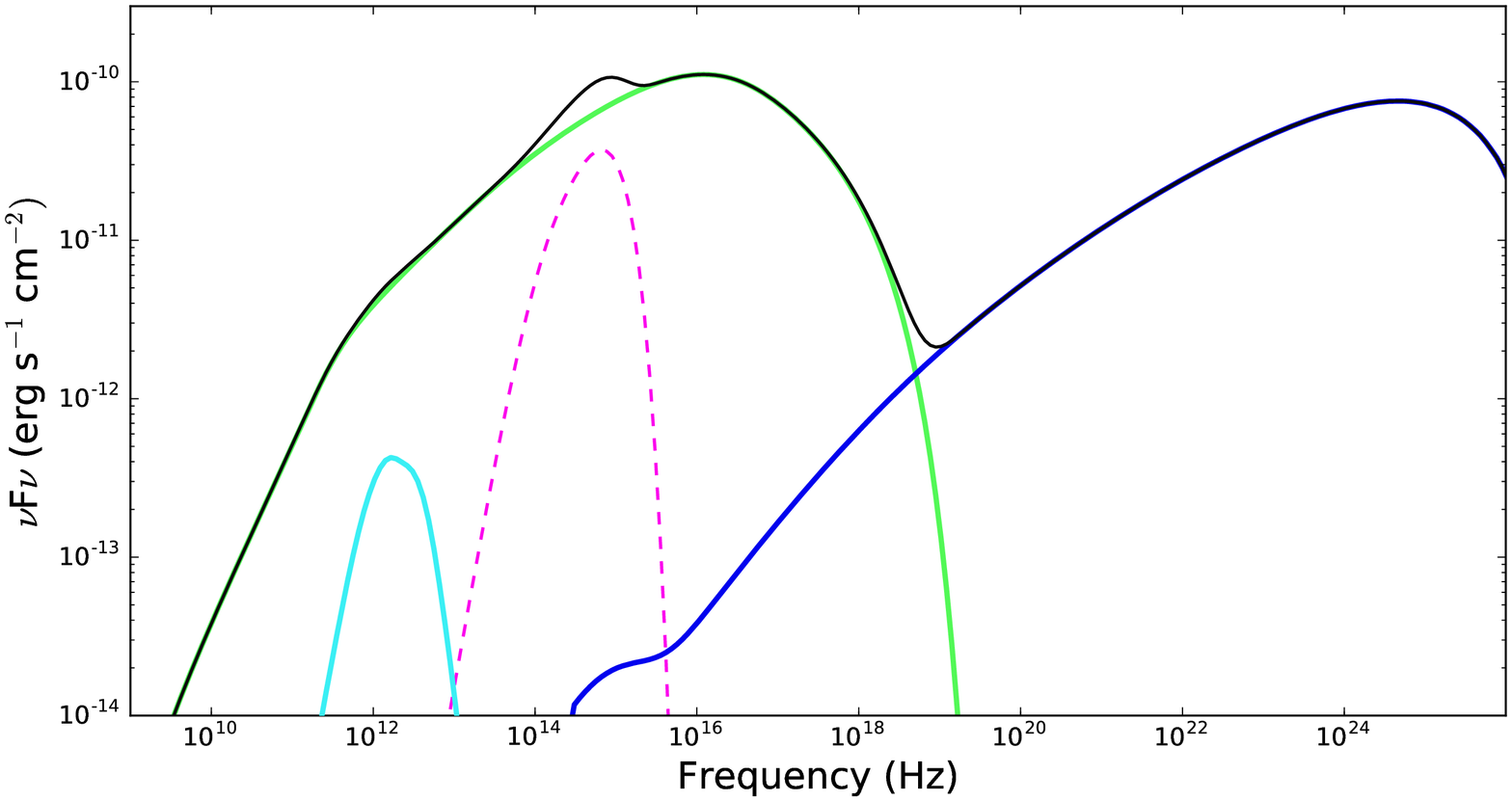}
\includegraphics[scale=.3]{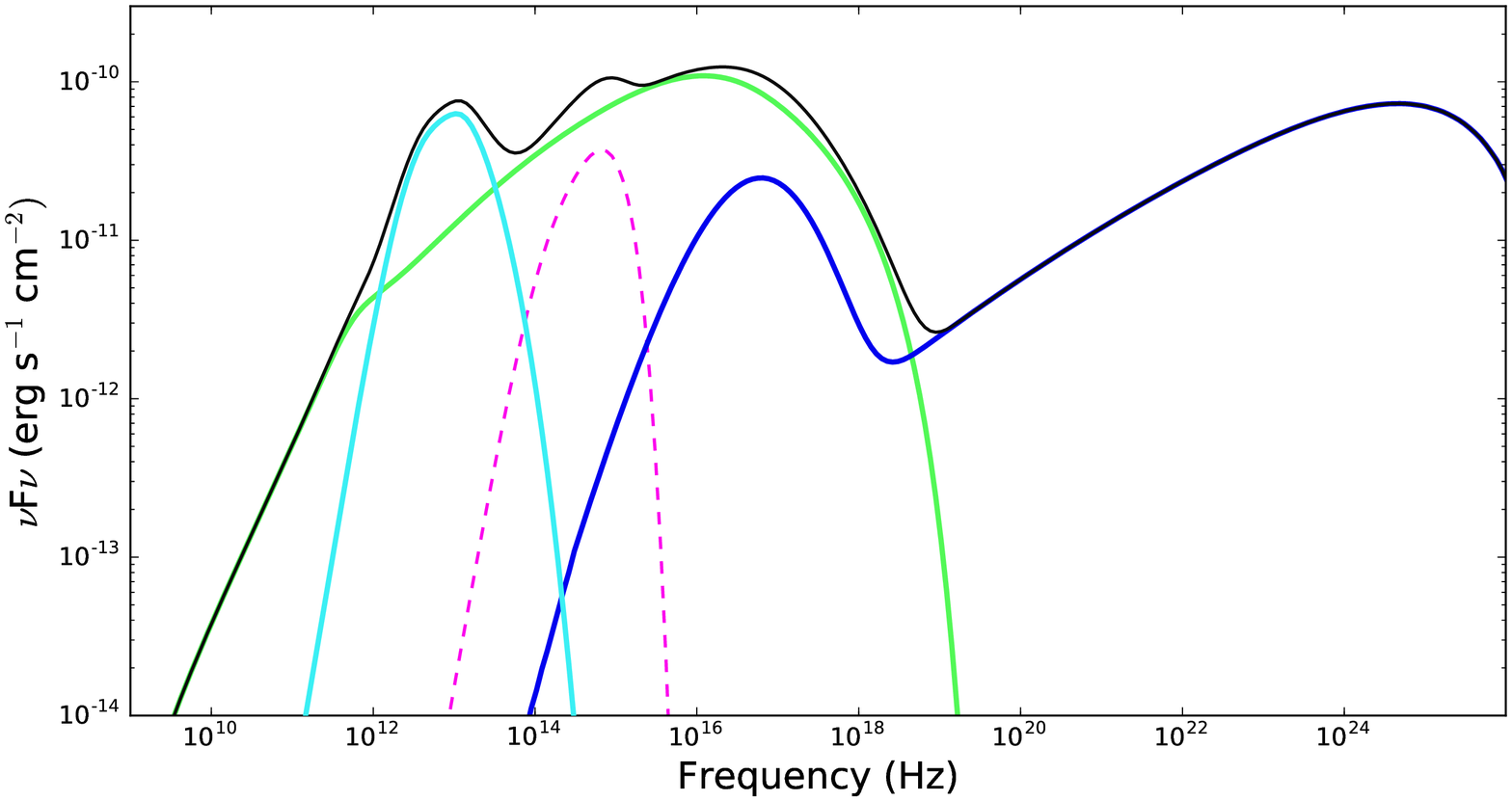}
\caption{
Changes in the SED as a function of the temperature in the nozzle, leaving the outer jet and non-thermal distribution unchanged; the left panel shows the emission for $\gamma_{\rm th} = 3$ and $f_{\rm heat} = 12$, the right panel for $\gamma_{\rm th} = 12$ and $f_{\rm heat} = 3$. The non-thermal synchrotron is shown in green, SSC in blue, the accretion disk in pink, the thermal synchrotron emission in cyan. If the nozzle temperature increases the thermal synchrotron emission results in a bump at mm/far-IR frequencies and brighter soft X-ray emission. 
}
\label{eltemp}
\end{figure*}

The typical alternative to SSC is to invoke external photon fields for Compton scattering, such as the BLR or torus. Despite our need for a disc contribution, which implies that in principle either of these mechanisms could be present in PKS\,2155$-$304, we find either EC scenario to be unlikely. The reason is the following: assume that the BLR/torus, if present, are hidden by the jet/accretion disc continuum, as one would expect in a BL Lac. With the standard distance scalings for the BLR and torus (e.g. \citeauthor{Ghisellini09} \citeyear{Ghisellini09}):
\begin{equation}
R_{\rm BLR} = 10^{17}L_{\rm d,45}^{1/2}\,\rm{cm}, \quad R_{\rm DT} = 2.5\cdot10^{18}L_{\rm d,45}^{1/2}\,\rm{cm},
\end{equation}
which for our disk parameters gives $R_{\rm BLR} \approx 400\,\rm{R_g}$ and $R_{\rm DT} \approx 10^{4}\,\rm{R_g}$ for the broad line region and torus, respectively. Assuming these scales are correct this immediately rules out the BLR, as it would lie closer to the BH than the dissipation region and therefore its seed photons would be strongly de-boosted in the co-moving frame of the jet. While a torus contribution may be present, dusty tori are generally not detected in FR\RN{1} sources (\citealt{vanderWolk10}, \citealt{Plotkin12}) thus making the presence of one in PKS\,2155$-$304 unlikely.

In \texttt{bljet} the leptons are described by a relativistic, thermal distribution (which for highly beamed sources does not contribute to the observed high-energy emission), with 10\% of the particles being channelled in a non-thermal tail responsible for the bulk of the emission; the thermal particles effectively act as a ``pool'' to replenish the non-thermal tail as it cools. We find that we require a relatively low particle acceleration efficiency (described by the parameter $f_{\rm sc}$), and that the temperature of the thermal ``pool'' has to increase significantly between the jet base and the outer jet regions. In particular, our \texttt{emcee} runs strongly rule out a scenario in which the temperature is the  same in the jet nozzle/corona and blazar zone ($f_{\rm heat} = 1$), requiring large amounts of heating instead ($f_{\rm heat} \gg 1$). We note that the amount of heating required is even higher if the leptons are initially non-relativistic ($T_{\rm e} < 511$ keV), as inferred by other corona models such as \texttt{nthcomp} (\citealt{Zdziarski96}, \citealt{Zycki99}). If instead the temperature in the nozzle is increased by a large amount (thus requiring lower heating far out in the jet), the thermal synchrotron emission from the inner jet regions results in a bright bump at mm/far-IR frequencies (and enhanced soft X-ray emission), as shown in Fig.~\ref{eltemp}. Such a spectral feature has never been observed in a blazar, and therefore we deem this scenario to be unphysical; as shown in Fig.~\ref{eltemp} our choice of $\gamma_{\rm th} = 3$ roughly corresponds to the highest allowed temperature in the nozzle/jet acceleration region. The need for large amounts of heating from the corona/jet base to the outer regions is consistent with the findings of \cite{Connors18}, who conducted a similar study of the black hole binary GX~339$-$4 using \texttt{agnjet} and who require a non relativistic plasma to model the X-ray spectra of the source. 

The required heating and low magnetization needed at the dissipation region favours shocks (which are only efficient as long as $\sigma \ll 1$, \citealt{Sironi11}, \citealt{Sironi13}, \citealt{Sironi15}) as the particle acceleration mechanism within the jet of in PKS\,2155$-$304. We can reproduce the long-term variability of the source (by long term here we mean the monthly/yearly periods during which each TANAMI SED was taken) purely by varying the slope, break and maximum energies of the non-thermal tail, suggesting that in the internal shock scenario the specific plasma conditions within the shocks change between epochs, leading to changes in the particle distribution.

Previous studies of BHBs and LLAGN with \texttt{agnjet} (e.g. \citealt{Connors17}, Connors et al. submitted, \citealt{Markoff05}, \citealt{Gallo07}, \citealt{Markoff07}, \citeyear{Markoff08}, \citealt{Maitra09}, \citealt{Plotkin15}, \citealt{Markoff15}, \citealt{Prieto16}) typically found a degeneracy between synchrotron-dominated and SSC-dominated scenarios to reproduce the X-ray emission of these sources. The key difference between the two is that synchrotron dominated fits require high acceleration efficiencies (corresponding to $f_{\rm sc} \approx 0.1$) in order to accelerate electrons at high enough energies to extend the non-thermal synchrotron emission up to the X-ray band, while this is not necessary for a purely SSC scenario. If we assume that the acceleration efficiency in BL Lacs jets is comparable to that of other sources, then our study favours the SSC case for BHBs and LLAGN (the different regimes for jet acceleration in the two models would not impact the inferred values of $f_{\rm sc}$). This result highlights the usefulness of studying jetted black hole sources as a whole to better constrain their properties.

\section{Conclusion}

In this work we have presented a new dynamical jet model for accreting black holes; the jet in the launching region is assumed to be a Poynting-dominated outflow which then accelerates by turning the magnetic flux into bulk kinetic energy. Energy is conserved during the acceleration process as long as the energy budget of the leptons is negligible compared to that of the protons. The treatment of radiation and particle acceleration is the same as the \texttt{agnjet} model by \cite{Markoff05}, \cite{Maitra09} and \citealt{Connors18} (submitted). 

As a benchmark for the model we fit six quasi-simultaneous, radio through $\gamma$-ray SEDs of the HSP BL Lac PKS\,2155$-$304; our modelling shows how even a very simple but physical treatment of a magnetically-accelerated jet is capable of linking a one-zone-like dissipation region with the launching mechanisms near the central engine, while keeping the number of free parameters (8-10) comparable to that of a one-zone model. Unlike one-zone models however, \texttt{bljet} also reproduces the radio data points, produced downstream in the jet away from the blazar zone. For the first time we have applied a joint fitting technique in order to break model degeneracies to a multi-wavelength dataset of a blazar. We find that the joint fit recovers parameters similar to those of an individual fit, and also discriminates much effectively between the various degenerate solutions allowed by single-epoch datasets. As a result, the parameter space of our joint fit is much simpler and most of the model's parameters can be estimated with reasonably small uncertainties.

The joint fit shows three main trends. First, we can model the long-term variability of the source with a steady-state jet in which the bulk properties of the outflow (geometry, magnetization, injected power) are unchanged and very well constrained by the joint fit, while the parameters of the radiating particles are free to vary. Second, in order to reproduce the observed $\gamma$-ray emission with SSC the jet has to be particle-dominated ($\sigma_{\rm diss} = 2.5^{+0.1}_{-0.2}\cdot 10^{-2}$) in the regions where the bulk of the jet's emission is produced. Third, the inferred energy budget for the jet ($N_{\rm j}= 0.90^{+0.06}_{-0.07}\cdot 10^{-2}\,L_{\rm Edd}$ for a $10^{9}\,M_{\odot}$ black hole) and the observed optical flux imply a contribution from the accretion disk in this band. The inferred accretion rate and jet power are found to be of the same order of magnitude; due to modelling uncertainties, we cannot estimate whether the jet power is higher than the accretion rate (implying a Blandford-Znajek origin for the jet) or not.

Despite the increasing quality of multi-wavelength data of jets in various sources, we are still far from a fully self-consistent model for jetted black holes. In order to capture the physics of the outflow such a model would have to include a more physical treatment of relativistic MHD (e.g. \citealt{Polko10}, \citeyear{Polko13}, \citeyear{Polko14}, \citealt{Ceccobello18}) but still be capable of producing spectra to be compared with observations. However, simpler approaches such as the one presented here are valuable for constraining the viable parameter space for more complex models.  We will cover these in future works. 

\section*{}
The authors of this manuscript would like to thank the suggestions of the anonymous referee for careful reading and useful comments. M. L. and S. M. are thankful for support from an NWO (Netherlands Organization for Scientific Research) VICI award, grant Nr. 639.043.513. F. K. acknowledges funding from the European Union's Horizon 2020 research and innovation program under grant agreement No. 653477. M. L. thanks Phil Uttley for insightful discussions on the variability of accretion flows. This research has made use of ISIS functions (ISISscripts) provided by ECAP/Remeis observatory and MIT (http://www.sternwarte.uni-erlangen.de/isis/).

\end{document}